\documentclass[12pt]{article}
\usepackage[dvips]{graphics}
\newlength{\titlesep}
\newlength{\authorsep}
\makeatletter

\def\fnum@figure{FIG.~\thefigure}

\newcounter{figureparent}
\@addtoreset{figure}{figureparent}

\newcounter{eqnparent}
\@addtoreset{equation}{eqnparent}

\renewcommand{\abstract}{\if@twocolumn
  \section*{Abstract}
  \else
  \begin{center}
    {\bf Abstract\vspace{-.5em}\vspace{0pt}}
  \end{center}
  \quotation
  \fi}
\renewcommand{\endabstract}{\if@twocolumn\else\endquotation\fi}

\newcommand{\thismonth}{\ifcase\month\or
 January\or February\or March\or April\or May\or June\or
 July\or August\or September\or October\or November\or December\fi
 \space \number\year}

\newcommand{\preprintnumber}[1]
{\begin{flushright}
  \begin{tabular}{l} #1 \end{tabular}
  \end{flushright}}

\makeatother

\newcommand{\Rn}[1]{{\uppercase\expandafter{\romannumeral#1}}}
\newcommand{\gsim}%
{\mathrel{\mbox{\raisebox{-1.0ex}%
{$\stackrel{\textstyle >}{\textstyle \sim}$}}}}
\newcommand{\lsim}%
{\mathrel{\mbox{\raisebox{-1.0ex}%
{$\stackrel{\textstyle <}{\textstyle \sim}$}}}}

\newcommand{\Journal}[4]{{#1} {\bf #2}, {#4} {(#3)}}

\newcommand{\plb}{\sl Phys.~Lett.~{\bf B}}

\newcommand{\pr}{\sl Phys.~Rev.}
\newcommand{\pra}{\sl Phys.~Rev.~{\bf A}}
\newcommand{\prd}{\sl Phys.~Rev.~{\bf D}}
\newcommand{\prl}{\sl Phys.~Rev.~Lett.}

\newcommand{\npb}{\sl Nucl.~Phys.~{\bf B}}

\newcommand{\epsfile}[1]{\relax}

\begin{document}
\baselineskip 18pt
\begin{titlepage}
\preprintnumber{
KEK Preprint 99-36\\
KEK-TH-623\\
}
\vspace*{\titlesep}
\begin{center}
{\LARGE\bf
 $\mu \rightarrow e \gamma$ and $\mu \rightarrow 3e$ processes with polarized muons and supersymmetric grand unified theories}
\\
\vspace*{\titlesep}
{$^{1,~2}$\large Yasuhiro  Okada}\footnote{yasuhiro.okada@kek.jp},\\
{$^{1,~3}$\large Ken-ichi Okumura}\footnote{ken-ichi.okumura@kek.jp}\\
and
{$^{1}$\large Yasuhiro  Shimizu}\footnote{yasuhiro.shimizu@kek.jp}\\
\vspace*{\authorsep}
{$^{1}$\it Theory Group, KEK, Tsukuba, Ibaraki, 305-0801 Japan }\\
{\it and}\\
{$^{2}$\it Department of Particle and Nuclear Physics and}\\
{$^{3}$\it Department of Accelerator Science,}\\
{\it The Graduate University of Advanced Studies, Tsukuba, Ibaraki, 305-0801 Japan}
\\
\end{center}
\vspace*{\titlesep}
\begin{abstract}
$\mu^{+} \rightarrow e^{+} \gamma$ and $\mu^{+} \rightarrow e^{+}e^{+}e^{-}$ processes are analyzed in detail with polarized muons in supersymmetric grand unified theories. 
We first present Dalitz plot distribution for $\mu^{+} \rightarrow e^{+}e^{+}e^{-}$ decay based on effective Lagrangian with general lepton-flavor-violating couplings and define various P- and T-odd asymmetries.
We calculate branching ratios and asymmetries in supersymmetric SU(5) and SO(10) models taking into account complex soft supersymmetry breaking terms.
Imposing constraints from experimental bounds on the electron, neutron and atomic electric dipole moments, we find that the T-odd asymmetry for $\mu^{+} \rightarrow e^{+}e^{+}e^{-}$ can be $15\%$ in the SU(5) case.
P-odd asymmetry with respect to muon polarization for $\mu^{+} \rightarrow e^{+} \gamma$  varies from $-20\%$ to $-100\%$ for the SO(10) model while it is $+100\%$ in the SU(5) case.
We also show that the P-odd asymmetries in $\mu^{+} \rightarrow e^{+}e^{+}e^{-}$ and the ratio of $\mu^{+} \rightarrow e^{+}e^{+}e^{-}$ and $\mu^{+} \rightarrow e^{+} \gamma$ branching fractions are useful to distinguish different models.
\end{abstract}
\end{titlepage}

\section{Introduction}

In order to explore physics beyond the standard model (SM), rare decay experiments can play a complementary role to direct search for new particles at high energy frontier.
Through forbidden or very suppressed processes within the minimal SM,
we may be able to obtain information on interaction at the energy scale not accessible by collider experiments.
Search for lepton flavor violation (LFV) is one of such windows to new physics. 

In recent years, LFV processes have received much attention because in the supersymmetric grand unified theory (SUSY GUT) the branching ratios for $\mu^{+}\rightarrow e^{+}\gamma$ and $\mu^{+}\rightarrow e^{+}e^{+}e^{-}$ and the $\mu$-$e$ conversion rate in a nucleus can reach just below present experimental values
 \cite{94Barbieri}-\cite{SO10}.
The present experimental upper bounds of these LFV processes are 
$B(\mu^{+} \rightarrow e^{+} \gamma) \leq 1.2 \times10^{-11}$ \cite{MEGA}, $B(\mu^{+} \rightarrow e^{+}e^{+}e^{-}) \leq 1.0\times 10^{-12}$\cite{mu3eex} and $\sigma(\mu^{-}$Ti$ \rightarrow e^{-}$Ti$)/\sigma(\mu^{-}$Ti$ \rightarrow capture) \leq 6.1\times10^{-13}$ \cite{conversionex}.
It is possible that future experiments will improve the sensitivity by two or three orders of magnitude below the current bounds \cite{LOI,Molzon:1998fp}.

In this paper we discuss the  $\mu^{+}\rightarrow e^{+}\gamma$ and $\mu^{+} \rightarrow e^{+}e^{+}e^{-}$ processes in SUSY GUT.
We focus on various asymmetries defined with the help of initial muon polarization.
Experimentally, polarized positive muons are available by the surface muon method because muons emitted from $\pi^{+}$'s stopped at target surface are $100\%$ polarized in the direction opposite to the muon momentum \cite{76Pifer}.
It is shown in Ref.\cite{96KunoOkada} that the muon polarization is useful to suppress the background processes in the $\mu^{+}\rightarrow e^{+}\gamma$ search.
As for the signal distribution of $\mu^{+}\rightarrow e^{+}\gamma$, the angular distribution with respect to the muon polarization can distinguish between $\mu^{+}\rightarrow e_{L}^{+}\gamma$ and $\mu^{+}\rightarrow e_{R}^{+}\gamma$. 
For $\mu^{+}\rightarrow e^{+}e^{+}e^{-}$, distribution in the Dalitz plot and various asymmetries defined with help of the muon polarization carry information on chirality and Lorentz structure of LFV couplings.  
In particular, we can define T-odd asymmetry which is sensitive to CP violation in LFV interactions \cite{77Treiman}. 
 In the previous paper \cite{letter} we pointed out that sizable T-odd asymmetry can occur in the SU(5) SUSY GUT when a CP violating phase is introduced in one of the soft SUSY breaking parameters, \it i.e. \rm the universal trilinear scalar coupling constant $A_{0}$.  
 The purpose of this paper is to give a model-independent framework for analyzing the $\mu^{+}\rightarrow e^{+}\gamma$ and $\mu^{+} \rightarrow e^{+}e^{+}e^{-}$ processes and investigate specific features of the SU(5) and SO(10) SUSY GUT focusing on the T-odd and other asymmetries. 
Detailed comparison of the T-odd asymmetry with the electron, neutron and Hg electric dipole moments (EDM) is also done introducing SUSY CP violating phases within the minimal supergravity (SUGRA) model.

 In section 2 we describe effective Lagrangian of the processes $\mu^{+} \rightarrow e^{+}\gamma$ and $\mu^{+} \rightarrow e^{+}e^{+}e^{-}$. 
We introduce a P-odd asymmetry for $\mu^{+}\rightarrow e^{+}\gamma$ and two types of P-odd asymmetries and a T-odd asymmetry for $\mu^{+} \rightarrow e^{+}e^{+}e^{-}$.  
 In section 3 we introduce the SU(5) and SO(10) SUSY GUT and briefly review how LFV processes arise in these theories.  
 In section 4 we present results of our numerical calculations.
We calculate the branching ratios and the asymmetries in the SUSY SU(5) and SO(10) models taking into account complex soft SUSY breaking terms under the constraints imposed by the EDM experiments.
We find that the T-odd asymmetry can be $15\%$ in the SU(5) case while it is less than $0.01\%$ in the SO(10) case.
We also show that the P-odd asymmetry for $\mu^{+}\rightarrow e^{+}\gamma$ varies $-20\%$~--~$-100\%$ for the SO(10) model and $100\%$ for the SU(5) case.
In the SU(5) case the P-odd asymmetries in $\mu^{+} \rightarrow e^{+}e^{+}e^{-}$ can reach $\pm30\%$ and the ratio of $\mu^{+}\rightarrow e^{+}\gamma$ and $\mu^{+}\rightarrow e^{+}e^{+}e^{-}$ branching fractions varies over SUSY parameter space.
On the contrary these asymmetries are smaller and the ratio of two branching fractions is almost constant in the SO(10) case.
In Appendices, useful formulas are listed.

\section{Phenomenology of the $\mu^{+} \rightarrow e^{+}\gamma$ and $\mu^{+} \rightarrow e^{+}e^{+}e^{-}$ processes}

 We begin with the effective Lagrangian for $\mu^{+}\rightarrow e^{+}\gamma$ and $\mu^{+} \rightarrow e^{+}e^{+}e^{-}$ processes. Using the electromagnetic gauge invariance and the Fierz rearrangement we can write without loss of generality:

\begin{eqnarray}
{\cal L} &=& -\frac{4G_F}{\sqrt{2}}\{  
        {m_{\mu }}{A_R}\overline{\mu_{R}}
        {{\sigma }^{\mu \nu}{e_L}{F_{\mu \nu}}}
       + {m_{\mu }}{A_L}\overline{\mu_{L}}
        {{\sigma }^{\mu \nu}{e_R}{F_{\mu \nu}}} \nonumber \\
    &&   +{g_1}(\overline{{{\mu }_R}}{e_L})
              (\overline{{e_R}}{e_L})
       + {g_2}(\overline{{{\mu }_L}}{e_R})
              (\overline{{e_L}}{e_R}) \nonumber \\
    &&   +{g_3}(\overline{{{\mu }_R}}{{\gamma }^{\mu }}{e_R})
              (\overline{{e_R}}{{\gamma }_{\mu }}{e_R})
       + {g_4}(\overline{{{\mu }_L}}{{\gamma }^{\mu }}{e_L})
              (\overline{{e_L}}{{\gamma }_{\mu }}{e_L})  \nonumber \\
    &&   +{g_5}(\overline{{{\mu }_R}}{{\gamma }^{\mu }}{e_R})
              (\overline{{e_L}}{{\gamma }_{\mu }}{e_L})
       + {g_6}(\overline{{{\mu }_L}}{{\gamma }^{\mu }}{e_L})
              (\overline{{e_R}}{{\gamma }_{\mu }}{e_R})
       +  h.c. \},
\label{eq:effective}
\end{eqnarray}
where $G_{F}$ is the Fermi coupling constant and $m_{\mu}$ is the muon mass.
The chirality projection is defined by the projection operators $P_{R}=\frac{1+\gamma_{5}}{2}$ and $P_{L}=\frac{1-\gamma_{5}}{2}$.
 $\sigma_{\mu\nu}$ is defined as $\sigma_{\mu\nu} = \frac{i}{2}[\gamma_{\mu},\gamma_{\nu}]$.  
$A_{L}$($A_{R}$) is the dimensionless photon-penguin coupling constant which contributes to $\mu^{+} \rightarrow e_{L}^{+}\gamma$ ($\mu^{+} \rightarrow e_{R}^{+}\gamma$).
These couplings also induce the $\mu^{+} \rightarrow e^{+}e^{+}e^{-}$ process. 
$g_{i}$'s $\; (i=1-6)$ are dimensionless four-fermion coupling constants which only contribute to $\mu^{+} \rightarrow e^{+}e^{+}e^{-}$.
$A_{L,R}$ and $g_{i}\; (i=1-6)$ are generally complex numbers and calculated based on a particular model with LFV interactions. 

The differential branching ratio for $\mu^{+} \rightarrow e^{+}\gamma$ is given by: 
~
\begin{eqnarray}
\frac{dB(\mu^{+} \rightarrow e^{+}\gamma)}{d\cos\theta} &=&
                192\pi^{2}\{|A_{L}|^{2}(1+\cos\theta)
                           +|A_{R}|^{2}(1-\cos\theta)\}
 \\
             &=& \frac{B(\mu^{+}\rightarrow e^{+}\gamma)}{2}
                 \{1+A(\mu^{+}\rightarrow e^{+}\gamma)P\cos\theta\},
\end{eqnarray}
where the total branching ratio for $\mu^{+}\rightarrow e^{+}\gamma$ ($B(\mu^{+}\rightarrow e^{+}\gamma)$) and the P-odd asymmetry ($A(\mu^{+}\rightarrow e^{+}\gamma)$) are defined as
\begin{eqnarray}
B(\mu^{+}\rightarrow e^{+}\gamma) &=& 384\pi^{2}(|A_{L}|^{2}+|A_{R}|^{2}),
 \\
A(\mu^{+}\rightarrow e^{+}\gamma) &=& \frac{|A_{L}|^{2}-|A_{R}|^{2}}{|A_{L}|^{2}+|A_{R}|^{2}}.
\end{eqnarray}
Here $P$ is the muon polarization and $\theta$ is the angle between the positron momentum and the polarization direction.

\begin{figure}[htbp]
  \begin{center}
    \leavevmode
     \scalebox{0.5}{\includegraphics{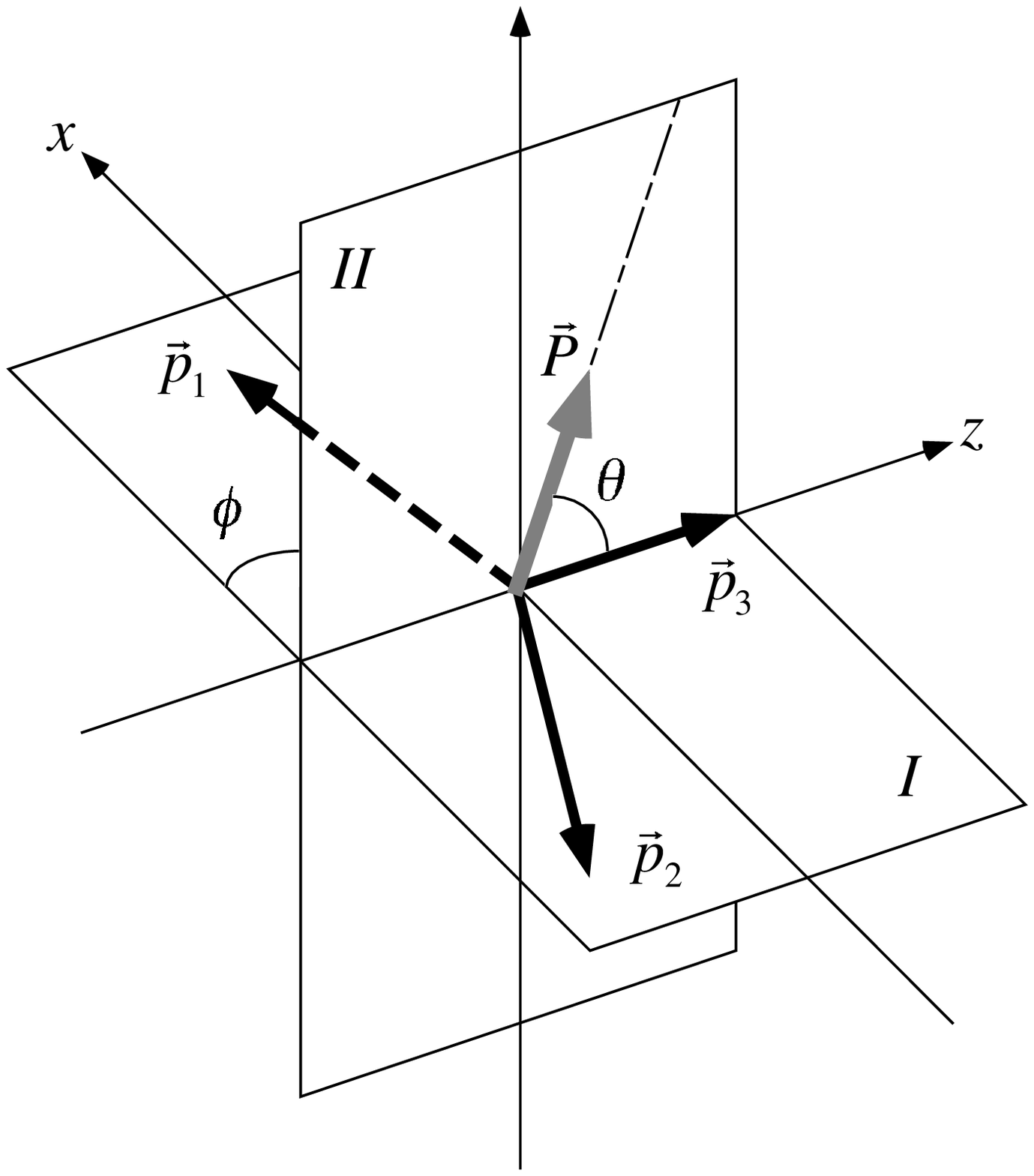}}
     \caption{Kinematics of the $\mu^{+} \rightarrow e^{+}e^{+}e^{-}$
       decay in the center-of-mass
      system of muon. The plane I represents the decay plane on which 
      the momentum vectors $\vec{p}_1$, $\vec{p}_2$, $\vec{p}_3$ lie,
      where $\vec{p}_1$ and $\vec{p}_2$ are  momenta of two $e^+$'s
      and  $\vec{p}_3$ is momentum of $e^-$ respectively.
      The plane II is the plane which the muon polarization vector
      $\vec{P}$ and $\vec{p}_3$ make. }
    \label{fig:kinematics}
  \end{center}
\end{figure}

\begin{figure}[htbp]
  \begin{center}
    \leavevmode
     \scalebox{0.8}{\includegraphics{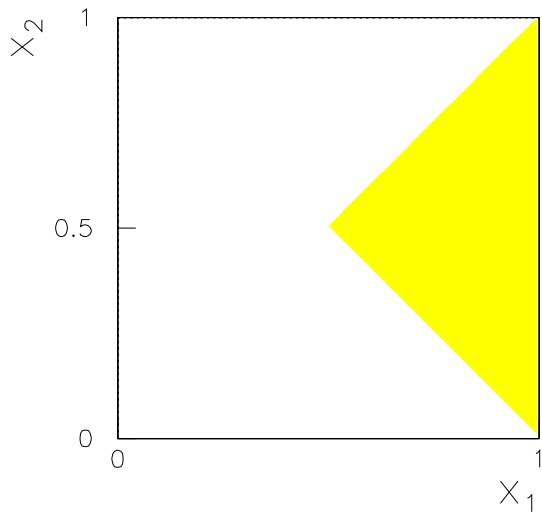}}
     \caption{Kinematical Region of the $\mu^{+}\rightarrow e^{+}e^{+}e^{-}$ decay
              in the center-of-mass system of muon. $x_1$ ($x_2$)
               represents a larger (smaller) energy of decay positrons
               normalized by $\frac{m_{\mu}}{2}$.
               A light shaded region is allowed.}
     \label{fig:dalitz}
  \end{center}
\end{figure}

 Kinematics of the $\mu^{+} \rightarrow e^{+}e^{+}e^{-}$ process with a polarized muon is determined by two energy variables of decay positrons and two angle variables which indicate the direction of the muon polarization with respect to the decay plane. 
 In Fig.~\ref{fig:kinematics} we take the $z$-axis as the direction of the decay electron momentum ($\vec{p_{3}}$) and the $z$-$x$ plane as the decay plane. 
Polar angles $(\theta,\varphi)$ $(0\leq \theta \leq\pi, 0\leq \varphi <2\pi)$ indicate the direction of the muon polarization $\vec{P}$.
 We take a convention that the decay positron having larger energy is named positron 1 and the other is positron 2 and $(p_1)_x \geq 0$. 
 We define the energy variables as $x_{1}=\frac{2E_{1}}{m_{\mu}}$ and $x_{2}=\frac{2E_{2}}{m_{\mu}}$ where $E_{1}$ and $E_{2}$ are the energy of the positron 1 and 2, respectively. 
 In this convention ($x_{1}$,$x_{2}$) represents one point of the Dalitz plot (Fig.~\ref{fig:dalitz}).
In our calculation we neglect the electron mass compared to the muon mass except for the total branching ratio.
In order to avoid logarithmic singularity we have to take into account the electron mass properly to evaluate the total branching ratio.

 Using the coupling constants in the Lagrangian in Eq.(\ref{eq:effective}) the differential branching ratio for $\mu^{+} \rightarrow e^{+}e^{+}e^{-}$ is written as follows:

\begin{eqnarray}
       \frac{dB}{dx_1 dx_2 d\cos \theta  d \varphi}&=&   
       \frac{3}{2\pi}[
       C_{1}\alpha_{1}(x_{1},x_{2})(1 + P\cos \theta)
          + C_{2}\alpha_{1}(x_{1},x_{2})(1 - P\cos \theta) \nonumber \\
         &+&C_{3}\{\alpha_{2}(x_{1},x_{2}) 
                   +P\beta_{1}(x_{1},x_{2})\cos \theta 
                   +P\gamma_{1}(x_{1},x_{2})\sin \theta \cos \varphi \} \nonumber \\
         &+&C_{4}\{\alpha_{2}(x_{1},x_{2}) 
                   -P\beta_{1}(x_{1},x_{2})\cos \theta 
                   -P\gamma_{1}(x_{1},x_{2})\sin \theta \cos \varphi \} \nonumber \\
         &+&C_{5}\{\alpha_{3}(x_{1},x_{2}) 
                   +P\beta_{2}(x_{1},x_{2})\cos \theta
                   +P\gamma_{2}(x_{1},x_{2})\sin \theta \cos \varphi \} \nonumber \\
         &+&C_{6}\{\alpha_{3}(x_{1},x_{2}) 
                   -P\beta_{2}(x_{1},x_{2})\cos \theta
                   -P\gamma_{2}(x_{1},x_{2})\sin \theta \cos \varphi \} \nonumber \\
         &+&C_{7}\{\alpha_{4}(x_{1},x_{2})(1-P\cos \theta) 
                  +P\gamma_{3}(x_{1},x_{2})\sin \theta \cos \varphi  \} \nonumber \\
         &+&C_{8}\{\alpha_{4}(x_{1},x_{2})(1+P\cos \theta) 
                  -P\gamma_{3}(x_{1},x_{2})\sin \theta \cos \varphi  \} \nonumber \\
         &+&C_{9}\{\alpha_{5}(x_{1},x_{2})(1+P\cos \theta)
                  -P\gamma_{4}(x_{1},x_{2})\sin \theta \cos \varphi\} \nonumber \\
         &+&C_{10}\{\alpha_{5}(x_{1},x_{2})(1-P\cos \theta)
                  +P\gamma_{4}(x_{1},x_{2})\sin \theta \cos \varphi\} \nonumber \\
         &+&C_{11}P\gamma_{3}(x_{1},x_{2})\sin \theta \sin \varphi
           -C_{12}P\gamma_{4}(x_{1},x_{2})\sin \theta \sin \varphi ], 
\label{eq:branching}
\end{eqnarray}
where $C_{i}$ are expressed by the coupling constants $g_{i}\; (i=1-6)$ and $A_{L,R}$ as: 		

\begin{eqnarray}
&& C_{1} = \frac{|g_{1}|^{2}}{16} + |g_{3}|^{2},~  
C_{2} = \frac{|g_{2}|^{2}}{16} + |g_{4}|^{2},~ \nonumber \\
&& C_{3} = |g_{5}|^{2},~ C_{4} = |g_{6}|^{2},C_{5} = |eA_{R}|^{2},~   
C_{6}   =   |eA_{L}|^{2}, \nonumber \\
&& C_{7}   =   Re(eA_{R}g_{4}^{*}),~ 			
C_{8}   =   Re(eA_{L}g_{3}^{*}),~
C_{9}   =   Re(eA_{R}g_{6}^{*}),~  			
C_{10}   =   Re(eA_{L}g_{5}^{*}),~ \nonumber \\ 			
&& C_{11}   =   Im(eA_{R}g_{4}^{*}+eA_{L}g_{3}^{*}),~
C_{12}   =   Im(eA_{R}g_{6}^{*}+eA_{L}g_{5}^{*}),
\end{eqnarray}
where $e(>0)$ is the positron charge and $P$ is the magnitude of the polarization vector.
Functions $\alpha_{i}$, $\beta_{i}$ and $\gamma_{i}$ are defined as: 

\begin{eqnarray}
\alpha_{1}(x_{1},x_{2}) & = & 8(2-x_{1}-x_{2})
                                         (x_{1}+x_{2}-1),
 \\
\alpha_{2}(x_{1},x_{2}) & = & 2\{x_{1}(1-x_{1})+x_{2}(1-x_{2})\},
 \\
\alpha_{3}(x_{1},x_{2}) & = & 8\{\frac{2x_{2}^{2}-2x_{2}+1}{1-x_{1}}
                                +\frac{2x_{1}^{2}-2x_{1}+1}{1-x_{2}}\}, 
 \\
\alpha_{4}(x_{1},x_{2}) & = & 32(x_{1}+x_{2}-1),
 \\
\alpha_{5}(x_{1},x_{2}) & = & 8(2-x_{1}-x_{2}),
 \\
\beta_{1}(x_{1},x_{2})  & = & 2\frac{ (x_{1}+x_{2})
                                    (x_{1}^{2}+x_{2}^{2})
                                    -3(x_{1}+x_{2})^{2}+6(x_{1}+x_{2})
                                    -4}{(2-x_{1}-x_{2})},
 \\
\beta_{2}(x_{1},x_{2})  & = & \frac{8}
                   {(1-x_{1})(1-x_{2})(2-x_{1}-x_{2})}\times \nonumber \\
                        &   &  \{
        2(x_{1}+x_{2})(x_{1}^{3}+x_{2}^{3})
     -4(x_{1}+x_{2})(2x_{1}^{2}+x_{1}x_{2}+2x_{2}^{2}) \nonumber \\
                        &   & +(19x_{1}^{2}+30x_{1}x_{2}+19x_{2}^{2})
     -12(2x_{1}+2x_{2}-1)
      \},
 \\       
\gamma_{1}(x_{1},x_{2}) & = & 4\frac{\sqrt{(1-x_{1})(1-x_{2})(x_{1}+x_{2}-1)}
                                     (x_{2}-x_{1})}{(2-x_{1}-x_{2})},
 \\
\gamma_{2}(x_{1},x_{2}) & = & 32\sqrt{\frac{(x_{1}+x_{2}-1)}
                                           {(1-x_{1})(1-x_{2})}}
                                \frac{(x_{1}+x_{2}-1)(x_{2}-x_{1})}
                                {(2-x_{1}-x_{2})},
 \\
\gamma_{3}(x_{1},x_{2}) & = & 16\sqrt{\frac{(x_{1}+x_{2}-1)}
                                           {(1-x_{1})(1-x_{2})}}
                                (x_{1}+x_{2}-1)(x_{2}-x_{1}),
 \\
\gamma_{4}(x_{1},x_{2}) & = & 8\sqrt{\frac{(x_{1}+x_{2}-1)}
                                           {(1-x_{1})(1-x_{2})}}
                                (2-x_{1}-x_{2})(x_{2}-x_{1}).
\end{eqnarray}

In Eq. (\ref{eq:branching}) 
there are three classes of terms: 
the first contribution arises from the four-fermion coupling constants ($C_{1-4}$) and the second from the photon-penguin coupling constants ($C_{5,6}$) and the third from interferences between the four-fermion couplings and the photon-penguin couplings ($C_{7 - 12}$).
The angular dependence with respect to the polarization direction is classified into four types, namely, terms proportional to (i) $1$, (ii) $\cos\theta$, (iii) $\sin\theta\cos\varphi$, and (iv) $\sin\theta\sin\varphi$.  
Under the parity operation (P), $\theta$, $\varphi$ transform as: 
\begin{eqnarray}
 \theta &\rightarrow& \pi - \theta, \nonumber \\
 \varphi &\rightarrow& \left \{
\begin{array}{ll}
\pi - \varphi & (0\leq \varphi < \pi) \\
3\pi - \varphi & (\pi\leq \varphi < 2\pi)
\end{array}
\right. ,
\end{eqnarray}
so that terms proportional to (ii) and (iii) are P-odd.
On the other hand the time reversal operation (T) induces the following
 transformation:
\begin{eqnarray}
&& \theta \rightarrow \theta, \; 
 \varphi\rightarrow 2\pi - \varphi.
\end{eqnarray}
Thus only terms proportional to $C_{11}$ and $C_{12}$ are T-odd quantities.
Notice that these terms are given by imaginary parts of interference terms between 
photon-penguin and four-fermion coupling constants. 
This means that effects of CP violation can be seen only through a phase difference between these two coupling constants.

It is convenient to define integrated asymmetries in order to separate four angular dependences, although in principle we can determine $C_{i}$ separately by fitting experimental data in full phase space. 
In the Dalitz plot, $\alpha_{3}$ and $\beta_{2}$ have a singularity as $\frac{1}{1-x_{1,2}}$ in the region near the kinematical boundary ($x_{1,2} \sim 1$). $\gamma_{2}$, $\gamma_{3}$ and $\gamma_{4}$ have a weaker singularity as $\frac{1}{\sqrt{1-x_{1,2}}}$. 
$\alpha_{3}$, $\beta_{2}$, and $\gamma_{2}$ arise as square of photon-penguin amplitudes whereas $\gamma_{3}$ and $\gamma_{4}$ from interferences between photon-penguin and four-fermion terms.
 On the contrary, contributions from square of the four-fermion coupling constants have no singularity on the edge and have a rather flat shape.  
These singular behaviors are cut off if we take into account the electron mass.
To show this behavior explicitly, we first integrate over smaller positron energy $x_2$ fixing the larger positron energy $x_1$ and define the following differential branching ratio and three types of asymmetries $a_{P_{1}}$,$a_{P_{2}}$ and $a_{T}$ as a function of the larger positron energy $x_1~(\frac{1}{2} \leq x_1 \leq 1 )$:

\begin{eqnarray}
\frac{dB(x_1)}{dx_1} & \equiv & \int^{x_1}_{1-x_1}dx_2
                       \int^{1}_{-1} d\cos\theta 
                       \int^{2\pi}_{0} d\varphi
                       \frac{dB}{dx_{1}dx_{2}d\cos\theta d\varphi}  \nonumber \\
              & = & 3
                    \{(C_{1}+C_{2})F_{1}(x_1)+(C_{3}+C_{4})F_{2}(x_1) \nonumber \\
              &   & +(C_{5}+C_{6})F_{3}(x_1)+(C_{7}+C_{8})F_{4}(x_1) \nonumber \\
              &   & +(C_{9}+C_{10})F_{5}(x_1)\},
 \\
a_{P_{1}}(x_1) & \equiv & \frac{1}{P\frac{dB(x_1)}{dx_1}}(
                       \int^{x_1}_{1-x_1}dx_2
                       \int^{1}_{0} d\cos\theta
                       \int^{2\pi}_{0} d\varphi
                       \frac{dB}{dx_{1}dx_{2}d\cos\theta d\varphi} \nonumber \\
              &   &   -\int^{x_1}_{1-x_1}dx_2
                       \int^{0}_{-1} d\cos\theta
                       \int^{2\pi}_{0} d\varphi
                       \frac{dB}{dx_{1}dx_{2}d\cos\theta d\varphi}
                       )  \nonumber \\
              & = &    \frac{3}{2}
                        \frac{1}{\frac{dB(x_1)}{dx_1}}\{
                        (C_{1}-C_{2})F_{1}(x_1)+(C_{3}-C_{4})G_{1}(x_1) \nonumber \\
              &   &    +(C_{5}-C_{6})G_{2}(x_1)-(C_{7}-C_{8})F_{4}(x_1) \nonumber \\
              &   &    +(C_{9}-C_{10})F_{5}(x_1)\},
 \\
a_{P_{2}}(x_1) & \equiv & \frac{-1}{P\frac{dB(x_1)}{dx_1}}
                       (
                       \int^{x_1}_{1-x_1}dx_2
                       \int^{1}_{-1} d\cos\theta
                       \int^{\frac{\pi}{2}}_{0} d\varphi
                       \frac{dB}{dx_{1}dx_{2}d\cos\theta d\varphi} \nonumber \\
              &   &   -\int^{x_1}_{1-x_1}dx_2
                       \int^{1}_{-1} d\cos\theta
                       \int^{\frac{3}{2}\pi}_{\frac{\pi}{2}} d\varphi
                       \frac{dB}{dx_{1}dx_{2}d\cos\theta d\varphi} \nonumber \\
              &   &   +\int^{x_1}_{1-x_1}dx_2
                       \int^{1}_{-1} d\cos\theta
                       \int^{2\pi}_{\frac{3}{2}\pi} d\varphi
                       \frac{dB}{dx_{1}dx_{2}d\cos\theta d\varphi}
                       ) \nonumber \\
              & = &    \frac{3}{2}
                       \frac{1}{\frac{dB(x_1)}{dx_1}}\{
                       (C_{3}-C_{4})H_{1}(x_1)+(C_{5}-C_{6})H_{2}(x_1)
                       \nonumber \\
              &   &   +(C_{7}-C_{8})H_{3}(x_1)-(C_{9}-C_{10})H_{4}(x_1)\},
 \\
a_{T}(x_1) & \equiv &    \frac{-1}{P\frac{dB(x_1)}{dx_1}}
                      (
                       \int^{x_1}_{1-x_1}dx_2
                       \int^{1}_{-1} d\cos\theta
                       \int^{\pi}_{0} d\varphi
                       \frac{dB}{dx_1 dx_2 d\cos\theta d\varphi} \nonumber \\
           &&          -\int^{x_1}_{1-x_1}dx_2
                       \int^{1}_{-1} d\cos\theta
                       \int^{2\pi}_{\pi} d\varphi
                       \frac{dB}{dx_1 dx_2 d\cos\theta d\varphi} 
                      ) \nonumber \\
              & = &    \frac{3}{2}\frac{1}{\frac{dB(x_1)}{dx_1}} 
                        \{C_{11}H_{3}(x_1)
                         -C_{12}H_{4}(x_1)
                        \}.
\end{eqnarray}
 In these formulas, $F_{i}$, $G_{i}$ and $H_{i}$ are functions of the variable $x_1$ and their analytic forms are found in Appendix \ref{sec:asym}. 
$\frac{dB(x_1)}{dx_1}$, $a_{P_{1}}(x_1)$, $a_{P_{2}}(x_1)$ and $a_{T}(x_1)$ are defined to extract terms (i)-(iv) with different angular dependences and $a_{T}(x_1)$ is the T-odd quantity.
In the above expression $F_3(x_1)$ in $\frac{dB(x_1)}{dx_1}$ and $G_2(x_1)$ in $a_{P_1}$ have $\frac{1}{1-x_1}$ singularity.
Introducing the cutoff $\delta$ for variable $x_1$ and integrating over $\frac{1}{2} \leq x_1 \leq 1-\delta$, we define the integrated branching ratio $B$ and three asymmetries $A_{P_{1}}$,$A_{P_{2}}$ and $A_{T}$. 
\begin{eqnarray}
B[\delta] & = & \int_{\frac{1}{2}}^{1-\delta} dx_1 \frac{dB(x_1)}{dx_1} \nonumber \\
          & = & 3
                 \{(C_{1}+C_{2})I_{1}[\delta]
                 +(C_{3}+C_{4})I_{2}[\delta]
                 +(C_{5}+C_{6})I_{3}[\delta] \nonumber \\
          &   &  +(C_{7}+C_{8})I_{4}[\delta] 
                 +(C_{9}+C_{10})I_{5}[\delta]\},
 \\
A_{P_{1}}[\delta] & = & \frac{1}{B[\delta]} \int_{\frac{1}{2}}^{1-\delta} dx_1
                a_{1}(x_1)\frac{dB}{dx_1}(x_1) \nonumber \\
          & = & \frac{3}{2B[\delta]} \{(C_{1}-C_{2})I_{1}[\delta]
                +(C_{3}-C_{4})J_{1}[\delta]+(C_{5}-C_{6})J_{2}[\delta] \nonumber \\
          &   & -(C_{7}-C_{8})I_{4}[\delta] 
                +(C_{9}-C_{10})I_{5}[\delta]\},
 \\
A_{P_{2}}[\delta] & = & \frac{1}{B[\delta]} \int_{\frac{1}{2}}^{1-\delta} dx_1
                a_{2}(x_1)\frac{dB}{dx_1}(x_1) \nonumber \\
          & = & \frac{3}{2B[\delta]} \{(C_{3}-C_{4})K_{1}[\delta]
                +(C_{5}-C_{6})K_{2}[\delta] 
                +(C_{7}-C_{8})K_{3}[\delta] \nonumber \\
          &   & -(C_{9}-C_{10})K_{4}[\delta]\},
 \\
A_{T}[\delta] & = & \frac{1}{B[\delta]} \int_{\frac{1}{2}}^{1-\delta} dx_1
                a_{3}(x_1)\frac{dB}{dx_1}(x_1) \nonumber \\
          & = & \frac{3}{2B[\delta]} \{C_{11}K_{3}[\delta]
                                        -C_{12}K_{4}[\delta]\}.
\end{eqnarray}
 $I_{i}$, $J_{i}$ and $K_{i}$ are functions of the cutoff $\delta$ and their analytic forms are also found in Appendix \ref{sec:asym}. 
Note that $I_{3}[\delta]$ and $J_{2}[\delta]$ have a logarithmic singularity at $\delta=0$.
Because of this logarithmic dependence, the terms $|A_L|^2$ and $|A_R|^2$ dominate over other terms in the branching ratio if coupling constants $eA_L$, $eA_R$ and $g_i$ have similar magnitudes. 
On the other hand the numerator of $A_T$ does not have a singular behavior so that $A_T$ itself is suppressed when we take very small $\delta$.
In the latter analysis of SUSY GUT cases we introduce the cutoff $\delta$ to optimize the T-odd asymmetry.

We have to take into account the electron mass properly to get precise value of total branching ratio.
If the photon-penguin contribution dominates the branching ratio, we can derive a model-independent relation between the two branching ratios \cite{98HisanoNomura}:

\begin{eqnarray}
\frac{B(\mu^{+} \rightarrow e^{+}e^{+}e^{-})}{B(\mu^{+} \rightarrow e^{+} \gamma)} &\simeq& \frac{\alpha}{3\pi}(\ln(\frac{m_{\mu}^{2}}{m_{e}^{2}})-\frac{11}{4}), \nonumber \\
  &\simeq& 0.0061,
\label{eq:ratio}
\end{eqnarray}
where $\alpha$ is the fine structure constant.
Neglecting the terms suppressed by $\frac{m_e}{m_{\mu}}$, the total branching ratio is , therefore, given by:

\begin{eqnarray}
B(\mu^{+}\rightarrow e^{+}e^{+}e^{-}) & = &  2(C_{1}+C_{2})
                 +(C_{3}+C_{4})
                 +32\{\log(\frac{m_{\mu}^2}{m_{e}^2})-\frac{11}{4}\}
                 (C_{5}+C_{6}) \nonumber \\
          &   &  +16(C_{7}+C_{8}) 
                 +8(C_{9}+C_{10}).
\end{eqnarray}

\section{SUSY GUT and LFV}

 In this section we introduce SU(5) and SO(10) SUSY GUT and discuss LFV processes.  
We assume that SUSY is broken explicitly at the Planck scale with soft SUSY breaking terms and that these terms have universal structure with respect to the flavor indices as suggested by the minimal SUGRA model. 
First, we discuss the LFV process in the SU(5) SUSY GUT and introduce the SO(10) SUSY GUT in the next subsection. 

\subsection{SU(5) SUSY GUT}
In the SU(5) SUSY GUT, we have three generations of $\mbox{\boldmath$10$} (T)$ and $\mbox{\boldmath$\overline{5}$} (\overline{F})$ representations of SU(5) as matter fields and $\mbox{\boldmath$5$} (H)$ and $\mbox{\boldmath$\overline{5}$} (\overline{H})$ representations of Higgs fields.
 The Yukawa superpotential and the soft SUSY breaking Lagrangian are written as follows:

\begin{eqnarray}
{\cal W}_{SU(5)} = \frac{1}{8}(y_{u})_{ij}T_{i}T_{j}H 
             + (y_{d})_{ij}\overline{F}_{i}T_{j}\overline{H},
\label{eq:Lagrangian su5}
\end{eqnarray}

\begin{eqnarray}
{\cal L}_{soft} & = & {-(m_{T}^{2}})_{ij}
                           \widetilde{T}_{i}^{\dag}\widetilde{T}_{j} 
                        - (m_{\overline{F}}^{2})_{ij}
                           \widetilde{\overline{F}}_{i}^{\dag}
                           \widetilde{\overline{F}}_{j}
                        -{m_{H}^{2}}
                          H^{\dag}H                        
                        -{m_{\overline{H}}^{2}}
                          \overline{H}^{\dag}\overline{H}
                          \nonumber \\
                   &   &-\{\frac{m_{0}}{8}(A_{u})_{ij}\widetilde{T}_{i}
                                          \widetilde{T}_{j}H
                          +m_{0}(A_{d})_{ij}
                        \widetilde{\overline{F}}_{i}\widetilde{T}_{j}\overline{H}
                        +\frac{1}{2}M_{5}\overline{\lambda_{5R}}{\lambda_{5L}}
                        +h.c. \}.
\end{eqnarray}
where $i,j$ are generation indices. 
$\widetilde{T}$, $\widetilde{\overline{F}}$ are scalar components of the superfields $T$, $\overline{F}$.

 At the Planck scale these soft SUSY breaking parameters satisfy flavor-blind universal conditions which are implied in the minimal SUGRA model:

\begin{eqnarray}
 & & m_{T}^{2} = m_{\overline{F}}^{2} = m_{0}^{2}\mbox{\boldmath$1$},\;
 m_{H}^{2} = m_{\overline{H}}^{2} = m_{0}^{2},\nonumber \\
 & & (A_{u})_{ij} = A_{0}(y_{u})_{ij},\;
 (A_{d})_{ij} = A_{0}(y_{d})_{ij}.
\end{eqnarray}
With these conditions the lepton and slepton mass matrices can be diagonalized simultaneously at the Planck scale, and therefore there is no LFV at this scale.
 However, these conditions receive corrections from the renormalization effect between the Planck scale and the GUT scale mainly due to the large top Yukawa coupling constant.
As a result the magnitude of the 3-3 element of the mass matrix for \boldmath$10\;$\unboldmath scalar fields becomes smaller than 1-1 and 2-2 elements.
 In the basis where $y_{u}$ is diagonalized at the Plank scale, the mass matrix for the \boldmath$10\;$\unboldmath scalar fields at the GUT scale is approximately given by:

\begin{eqnarray}
m_{T}^{2} &\simeq& \left(
\begin{array}{ccc}
m^{2} &       & \\
      & m^{2} & \\
      &       & m^{2}+\Delta m^{2}
\end{array}
\right), \nonumber \\
\Delta m^{2} &\simeq& -\frac{3}{8\pi^{2}}|(y_{u})_{33}|^{2}m_{0}^{2}(3 + |A_{0}|^{2})\ln(\frac{M_{P}}{M_{G}}), 
\label{massdifference}
\end{eqnarray}
where $M_{P}$ and $M_{G}$ denote the reduced Planck mass ($\sim2\times10^{18}$GeV)and the GUT scale ($\sim2\times10^{16}$GeV).
 This correction amounts to about $50\%$ of their original values and the lepton and slepton mass matrices are no longer diagonalized simultaneously.
This becomes a source of LFV which could induce observable effects in $\mu^{+} \rightarrow e^{+} \gamma$\cite{94Barbieri}. 

 The SU(5) symmetry is broken to the SM groups at the GUT scale, and after integrating out heavy fields the effective theory becomes the minimal supersymmetric standard model (MSSM). The superpotential and the soft SUSY breaking Lagrangian for the MSSM are written as follows:  
\begin{eqnarray}
{\cal W}_{MSSM} &=& \epsilon^{\alpha\beta}{y_{e}}_{ij}
               H_{1\alpha} E_{i}^{c} L_{j\beta}
             +\epsilon^{\alpha\beta}{y_{d}}_{ij}
               H_{1\alpha} D_{i}^{c} Q_{j\beta} \nonumber \\
           &&  +\epsilon^{\alpha\beta}{y_{u}}_{ij}
               H_{2\alpha} U_{i}^{c} Q_{j\beta}
             +\epsilon^{\alpha\beta}\mu H_{1\alpha}H_{2\beta},
\label{eq:Lagrangian MSSM}
\end{eqnarray}
\begin{eqnarray}
{\cal L}_{ soft} &=&  -(m^{2}_{E})_{ij}\widetilde{E}_{i}^{*}
                                            \widetilde{E}_{j}
                         -(m^{2}_{L})_{ij}\widetilde{L_{i}}^{*}
                                            \widetilde{L_{j}}
                         -(m^{2}_{D})_{ij}\widetilde{D}_{i}^{*}
                                            \widetilde{D}_{j} \nonumber \\
                     & & -(m^{2}_{U})_{ij}\widetilde{U}_{i}^{*}
                                            \widetilde{U}_{j}  
                         -(m^{2}_{Q})_{ij}\widetilde{Q}_{i}^{*}
                                            \widetilde{Q}_{i}
                         -m^{2}_{H_{1}}H_{1}^{\dag}H_{1}
                         -m^{2}_{H_{2}}H_{2}^{\dag}H_{2}
                         \nonumber \\
                     & & -[m_{0}(A_{e})_{ij}
                           \epsilon^{\alpha\beta}H_{1\alpha}
                          \widetilde{E}_{i}^{*}\widetilde{L}_{j\beta}
                         +m_{0}(A_{d})_{ij}
                           \epsilon^{\alpha\beta}H_{1\alpha}
                          \widetilde{D}_{i}^{*}\widetilde{Q}_{j\beta} \nonumber \\
                     & & +m_{0}(A_{u})_{ij}
                           \epsilon^{\alpha\beta}H_{2\alpha}
                          \widetilde{U}_{i}^{*}\widetilde{Q}_{j\beta}
                         +\epsilon^{\alpha\beta}
                          \mu B H_{1\alpha}H_{2\beta} \nonumber \\ 
                     & & +\frac{1}{2}M_{1}\overline{\widetilde{B_{R}}}
                                          \widetilde{B_{L}}
                         +\frac{1}{2}M_{2}\overline{\widetilde{W_{R}}}
                                          \widetilde{W_{L}}
                         +\frac{1}{2}M_{3}\overline{\widetilde{G_{R}}}
                                          \widetilde{G_{L}}
                         +h.c. ].
\label{eq:Soft Breaking MSSM}
\end{eqnarray}
In this formula $\epsilon^{\alpha\beta}$ is defined as $\epsilon^{11}=\epsilon^{22}=0$, $\epsilon^{12}=-\epsilon^{21}=1$.
At the GUT scale these parameters satisfy the GUT relations:
\begin{eqnarray}
 && y_{e} = y_{d}^{T} , \label{eq:GUT Yukawa relation}\\
 && A_{e} = A_{d}^{T} , \\ 
 && m_{E}^{2T} = m_{U}^{2T} = m_{Q}^{2} = m_{T}^{2},\;
 m_{L}^{2} = m_{D}^{2} = m_{\overline{F}}^{2},\nonumber\\
 && m_{H_{1}}^{2} = m_{\overline{H}}^{2},\;
 m_{H_{2}}^{2} = m_{H}^{2},\; M_{1} = M_{2} = M_{3}  =  M_{5}.
\label{GUTrelation}
\end{eqnarray}
 In the basis where $y_{u}$ is diagonalized at the Planck scale, $y_{u}$ at the GUT scale still approximately remains diagonal.
In this basis, $y_{e}$ is diagonalized in the following way:

\begin{eqnarray}
 V_{R}y_{e}V_{L}^{\dag} &=& diagonal,
\end{eqnarray}
where $V_{L}$ and $V_{R}$ are unitary matrices and using Eq.(\ref{eq:GUT Yukawa relation}) $V_{R}$ is given by:
\begin{eqnarray}
(V_{R})_{ij} &=& (V_{KM}^{0})_{ji}, 
\end{eqnarray}
where $V_{KM}^{0}$ is the Kobayashi-Maskawa (KM) matrix at the GUT scale.

It is useful to make unitary transformations on $E_{i}$ and $L_{j}$ to go to the basis where $y_{e}$ is diagonalized at the GUT scale.
In the new basis the off-diagonal element of $m_{E}^{2}$ is given by:

\begin{eqnarray}
(m_{E}^{2})_{ij} \simeq -\frac{3}{8\pi^{2}}(V_{KM}^{0})_{3i}(V_{KM}^{0})_{3j}^{*}|(y_{u})_{33}|^{2}m_{0}^{2}(3 + |A_{0}|^{2})\ln(\frac{M_{P}}{M_{G}}).
\label{eq:m_R su5} 
\end{eqnarray}
The off-diagonal element of the slepton mass matrix becomes a source of LFV.

In the actual numerical analysis, we solved the MSSM renormalization group equation from the GUT scale to the electroweak scale and determine the masses and mixings for SUSY particles.
We also require the electroweak symmetry breaking occurs properly to give the correct $Z$-boson mass.
From the MSSM Lagrangian at the electroweak scale we can derive the LFV coupling constants $A_{L,R}$ and $g_{1-6}$ through 1-loop diagrams involving slepton, gaugino and higgsino.
The complete formulas are given in Appendix  \ref{sec:couplings}.

In the SU(5) model, only the right-handed slepton mass matrix can develop off-diagonal terms if the ratio of vacuum expectation values of two Higgs fields ($\tan \beta=\frac{\langle H_{2}^{0} \rangle}{\langle H_{1}^{0} \rangle}$) is not very large.
In such a case only $A_{L}$, $g_{3}$ and $g_{5}$ have sizable contributions.
Restricting to small or moderate $\tan\beta$ cases, 
all effective coupling constants are proportional to the product of the KM matrix element $\lambda_{\tau}=(V_{KM}^{0})_{32}(V_{KM}^{0})_{31}^{*}$ since the LFV transition occurs through $(m_{E}^2)_{21}$ or $(m_{E}^2)_{32}^*(m_{E}^2)_{31}$.
This situation does not change even if we take into account the LFV transition due to the left-right mixing of the slepton mass matrix.  
This means that the CP violating phase of Yukawa coupling constants cannot make a phase difference between $A_{L}$ and $g_{3}$, or $A_{L}$ and $g_{5}$, and therefore the T-odd asymmetry $A_{T}$ cannot appear from this source.   

There is another important source of CP violating phases in soft SUSY breaking terms.
Within the SUGRA model, we can introduce four phases: phases of $M_{5}$, $A_{0}$, $B$ and $\mu$, but not all of them are physically independent.
By field redefinition, we can take the phases of $A_{0}$ and $\mu$ as independent phases.
If we take into account these phases, $A_{T}$ can be generated. Since these phases also induce the electron, neutron and Hg EDMs\cite{EDM,Falk:1999tm},
we take into account these EDM constraints to obtain allowed region of SUSY phases.

Up to now we consider that the Yukawa coupling constants are given by Eq. (\ref{eq:Lagrangian su5}), so that the lepton and down-type quark Yukawa coupling constants are related at the GUT scale by Eq. (\ref{eq:GUT Yukawa relation}).
On the other hand, it is known that this relation does not reproduce realistic mass relations for charged leptons and down-type quarks in the first and second generations.
It is therefore important to study how the prediction for LFV processes  depends on details of the origin of the Yukawa coupling constant in the MSSM Lagrangian.
One way to generate realistic mass matrix is to introduce higher dimensional operators in the SU(5) superpotential.
Once this is done the simple relationship  between the charged lepton and down-type quark Yukawa coupling constants does not hold.
Although the effect of higher dimensional operators is suppressed by $O\left(\frac{M_{G}}{M_{P}}\right)$,
masses and mixings for the first and second generations can receive large corrections to the GUT relation.
If $\tan \beta$ is not very large, LFV is still induced only for the right-handed slepton sector and Eq. (\ref{eq:m_R su5}) holds with replacement of $V_{KM}^{0}$ by $V_{R}^{T}$ which is not necessary related to the KM matrix elements. 
In the followings, therefore, we treat $\lambda_{\tau}$ as a free parameter.
Since the $\mu^{+} \rightarrow e^{+} \gamma$ and the $\mu^{+} \rightarrow e^{+}e^{+}e^{-}$ branching ratios are proportional to $|\lambda_{\tau}|^{2}$, we present these branching ratios divided by $|\lambda_{\tau}|^{2}$.
If $\tan \beta$ is as large as 30, the bottom Yukawa coupling constant can induce the LFV in the left-handed slepton sector.
In such a case, if we include the effect of higher dimensional operators at the GUT scale, there are photon-penguin diagrams which are proportional to $m_{\tau}$ and these contributions tend to dominate over other contributions as shown in \cite{98HisanoNomuraOkada}. 
Because the LFV branching ratios depend on many unknown parameters in such a case, we do not consider this possibility here.

\subsection{SO(10) SUSY GUT}
In the minimal SO(10) model, we assume three generations of \boldmath$16\;$\unboldmath representation matter fields ($\Psi_{i}$) and two \boldmath$10\;$\unboldmath representation Higgs fields ($\Phi_{u}$,$\Phi_{d}$) of SO(10). 
The Yukawa superpotential and the soft SUSY breaking Lagrangian are written as follows:

\begin{eqnarray}
 {\cal W}_{ SO(10)} &=& \frac{1}{2}(y_{u})_{ij}\Psi_{i}\Phi_{u}\Psi_{j} 
             + \frac{1}{2}(y_{d})_{ij}\Psi_{i}\Phi_{d}\Psi_{j}, 
\end{eqnarray}

\begin{eqnarray}
{\cal L}_{soft} & = & {-(m_{\Psi}^{2}})_{ij}
                           \widetilde{\Psi}_{i}^{\dag}\widetilde{\Psi}_{j} 
                        -{m_{\Phi_{u}}^{2}}
                          \Phi_{u}^{\dag}\Phi_{u}                        
                        -{m_{\Phi_{d}}^{2}}
                          \Phi_{d}^{\dag}\Phi_{d}
                          \nonumber \\
                   &   &-\{\frac{m_{0}}{2}(A_{u})_{ij}\widetilde{\Psi}_{i}
                                          \Phi_{u}\widetilde{\Psi}_{j}
                          +\frac{m_{0}}{2}(A_{d})_{ij}
                          \widetilde{\Psi}_{i}\Phi_{d}\widetilde{\Psi}_{j}
                        +\frac{1}{2}M_{10}\overline{\lambda_{10R}}{\lambda_{10L}}
                  \nonumber \\
                  &   &      +h.c. \}.
\end{eqnarray}
At the Planck scale, we have the universal boundary conditions:

\begin{eqnarray}
 & & m_{\Psi}^{2} = m_{0}^{2}\mbox{\boldmath$1$},\;
 m_{\Phi_{u}}^{2} = m_{\Phi_{d}}^{2} = m_{0}^{2},\; 
 (A_{u})_{ij} = A_{0}(y_{u})_{ij},\;
 (A_{d})_{ij} = A_{0}(y_{d})_{ij}.
\end{eqnarray}
 In contrast with the SU(5) SUSY GUT, all matter fields are unified in a single representation $\Psi$ of SO(10) and masses of all squarks and sleptons of the third generation receive a large correction due to the renormalization effect by the top Yukawa coupling constant.
 In the $y_{u}$-diagonalized basis, difference between the mass of the third generation sfermion and that of the first and second generation is given by: 

\begin{eqnarray}
\Delta m_{\Psi}^{2} &\simeq& -\frac{5}{8\pi^{2}}|(y_{u})_{33}|^{2}m_{0}^{2}(3 + |A_{0}|^{2})\ln(\frac{M_{P}}{M_{G}}). 
\label{massdifference2}
\end{eqnarray}
At the GUT scale, the initial conditions for the parameters of MSSM Lagrangian in Eqs. (\ref{eq:Lagrangian MSSM}) and (\ref{eq:Soft Breaking MSSM}) at the GUT scale are written as follows:

\begin{eqnarray}
&& y_{e} = y_{d} ,\\
&& A_{e} = A_{d} ,\\
&& m_{E}^{2} = m_{L}^{2} = m_{D}^{2} = m_{U}^{2} = m_{Q}^{2} = m_{\Psi}^{2}, \nonumber \\
&& m_{H_{1}}^{2} = m_{\Phi_{d}}^{2},\;
m_{H_{2}}^{2}  =  m_{\Phi_{u}}^{2},\;
 M_{1} = M_{2} = M_{3} =  M_{10},
\label{GUTrelation2}
\end{eqnarray}
where the symmetric matrix $y_{e}$ can be expressed as:
\begin{eqnarray}
 y_{e} &=& U^{T}P\hat{y}_{e}U, \nonumber \\
 P &=& \left(
\begin{array}{ccc}
e^{i\phi_{1}} & & \\
              & e^{i\phi_{2}} & \\
              &               & e^{i\phi_{3}} \\
\end{array}
\right),
\label{eq:GUT phases}
\end{eqnarray}
where $\hat{y}_{e}$ is a real diagonal matrix,
and therefore the unitary matrix $U$ is related to the KM matrix at the GUT scale as:
\begin{eqnarray}
 U &=& V_{KM}^{0 \dag}.
\end{eqnarray} 
If we go to the $y_{e}$-diagonalized basis at the GUT scale, the off-diagonal elements of slepton mass matrices become as follows:

\begin{eqnarray}
(m_{E}^{2})_{ij} &\simeq& -\frac{5}{8\pi^{2}}e^{-i(\phi_{i}-\phi_{j})}(V_{KM}^{0})_{3i}(V_{KM}^{0})_{3j}^{*}|(y_{u})_{33}|^{2}m_{0}^{2}(3 + |A_{0}|^{2})\ln(\frac{M_{P}}{M_{G}}),
 \\
(m_{L}^{2})_{ij} &\simeq& -\frac{5}{8\pi^{2}}(V_{KM}^{0})_{3i}^{*}(V_{KM}^{0})_{3j}|(y_{u})_{33}|^{2}m_{0}^{2}(3 + |A_{0}|^{2})\ln(\frac{M_{P}}{M_{G}}). 
\end{eqnarray}
Since the left-handed slepton also has the LFV effect in the case of the SO(10) SUSY GUT, there are dominant photon-penguin diagrams which are proportional to $m_{\tau}$ in the slepton left-right mixing as discussed in \cite{95Barbieri} .

In addition to the KM phase, there are two physical phases in Eq. (\ref{eq:GUT phases}) up to a overall phase.
A combination of these phases and the KM phase is responsible to the electron EDM\cite{95DimopoulousHall},\cite{95Barbieri}.
If the photon-penguin diagram proportional to $m_{\tau}$ dominates in the $\mu^+ \rightarrow e^+\gamma$ amplitude, there is a simple relation between the electron EDM and the $\mu^+ \rightarrow e^+\gamma$ branching ratio \cite{95Barbieri}.
Defining a phase as
\begin{equation}
Im[e^{i(\phi_3-\phi_1)}\{(V_{KM}^0)_{31}(V_{KM}^{0})_{33}^*\}^2] = |(V_{KM}^0)_{31}(V_{KM}^{0})_{33}^*|^2\sin\phi,
\label{eq:relative phase}
\end{equation}
the relation is given by:
\begin{equation}
|d_e|=1.3\sqrt{\frac{B(\mu\rightarrow e\gamma)}{10^{-12}}}|\sin\phi|~~(10^{-27}e\cdot cm).
\label{eq:EDM relation}
\end{equation}   
Later we see that the diagram proportional to $m_{\tau}$ does not necessarily dominate over other diagrams.
In such a case the above relation does not hold.
\section{Results of numerical calculations}

 We present results of our numerical analysis on $\mu^{+} \rightarrow e^{+}\gamma$ and $\mu^{+} \rightarrow e^{+}e^{+}e^{-}$ processes for the SU(5) and SO(10) SUSY GUT.
We also calculate the electron, neutron and Hg EDMs as constraints on the CP violating phases of the soft SUSY breaking terms.
Following the procedure discussed in the previous section, we solve the renormalization group equations with the universal condition for the SUSY breaking terms at the Plank scale.
Though the approximate formulas for the slepton mass difference are given in the previous section to explain qualitative features, we solve the renormalization group equations from the Planck scale to the electroweak scale numerically taking into account the full flavor-mixing matrix for fermions and sfermions.
To determine allowed range of SUSY parameter space we use the results of various SUSY particle searches at LEP and Tevatron and the branching ratio $B(b \rightarrow s\gamma)$.
The details on these constraints are described in Ref. \cite{98GotoShimizu}\footnote{The branching ratio $B(b \rightarrow s\gamma)$ is updated as $2.0\times10^{-4}<B(B \rightarrow X_{s}\gamma)<4.5\times10^{-4}$ \cite{98CLEO}}.
We take the top quark mass as $m_{t} = 175$ GeV. 
Because we calculate the LFV branching ratios divided by $|\lambda_{\tau}|^{2}$, the result is almost independent of the KM matrix elements.
For definiteness, we use the input parameters of the KM matrix elements as $|(V_{KM})_{cb}|=0.041$, $|(V_{KM})_{td}|=0.006$ and $|(V_{KM})_{us}|=0.22$.
Requiring the radiative electroweak symmetry breaking the free parameters of the SUGRA model can be taken as $\tan\beta$, $M_{2}$, $m_{0}$, $|A_{0}|$ and the phase of $A_{0}$ ($\theta_{A_{0}}$) and that of $\mu$ ($\theta_{\mu}$).

\subsection{SU(5) GUT}
\begin{figure}[htbp]
  \begin{center}
    \leavevmode
    \scalebox{.8}{\includegraphics{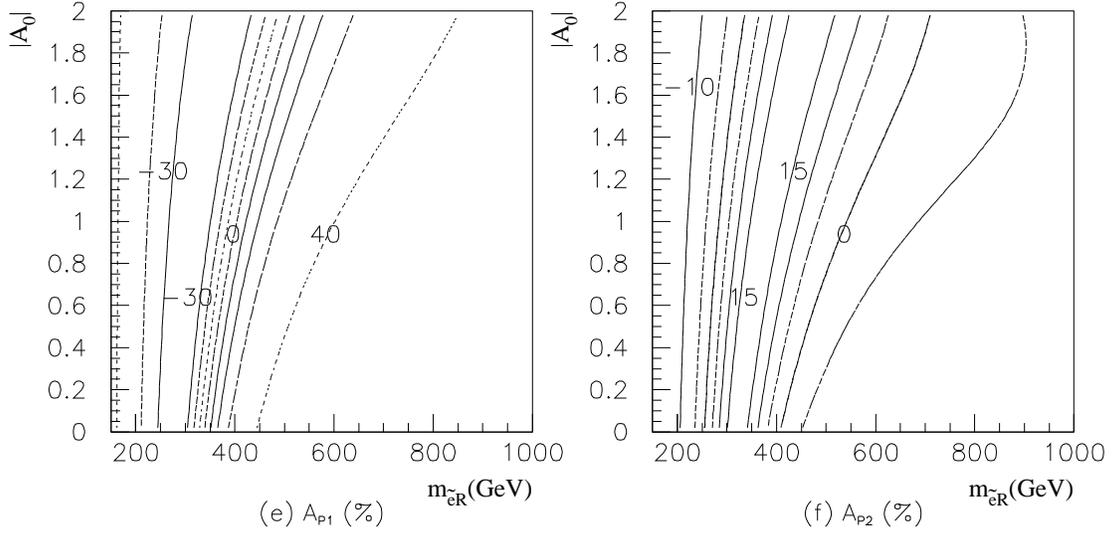}}
    \caption{The observables in the SU(5) model
             without the SUSY CP violating phases 
             in the $m_{\tilde{e}_R}$-$|A_0|$ plane. 
             We fix the SUSY parameters as $\tan\beta = 3$, 
             $M_2 = 150$ GeV and $\mu > 0$
              and the top quark mass as 175 GeV.
             (a) Branching ratio for $\mu^{+} \rightarrow e^{+} \gamma$
              normalized by $|\lambda_\tau|^2$$\equiv$ 
             $|(V_R)_{23}(V_R)^\ast_{13}|^2$.
             (b) Branching ratio for $\mu^{+} \rightarrow e^{+}e^{+}e^{-}$
              normalized by $|\lambda_\tau|^2$. 
             (c) The ratio of two branching fractions 
             $\frac{B(\mu\rightarrow 3e)}
             {B(\mu\rightarrow e\gamma)}$.  
             (d) The P-odd asymmetry for $\mu^{+} \rightarrow e^{+} \gamma$.
             (e) The P-odd asymmetries $A_{P_1}$ 
             for $\mu^{+} \rightarrow e^{+}e^{+}e^{-}$.
             (f) The P-odd asymmetries $A_{P_2}$ 
             for $\mu^{+} \rightarrow e^{+}e^{+}e^{-}$. 
             The cut-off parameter $\delta$
              is taken to be 0.02.}
    \label{fig:su5_1}
  \end{center}
\end{figure}

Let us first discuss the case without the CP violating phases in the SU(5) GUT.
In Fig.~\ref{fig:su5_1} we present the following quantities
\begin{eqnarray}
\label{eq:data set}
&& \frac{B(\mu^{+}\rightarrow e^{+}\gamma)}{|\lambda_{\tau}|^{2}},~
\frac{B(\mu^{+}\rightarrow e^{+}e^{+}e^{-})}{|\lambda_{\tau}|^{2}},~
\frac{B(\mu^{+}\rightarrow e^{+}e^{+}e^{-})}
{B(\mu^{+}\rightarrow e^{+}\gamma)},~\nonumber \\
&& A(\mu^{+}\rightarrow e^{+}\gamma),~
A_{P_{1}}~,
A_{P_{2}},
\end{eqnarray}
in the plane of $m_{\tilde{e}_{R}}$ and $|A_{0}|$ for $\tan\beta = 3$,
 $M_{2} = 150$GeV, $\theta_{A_{0}}=\theta_{\mu}=0$.
Here $\lambda_{\tau}$ is defined by the mixing matrix which diagonalizes the right-handed slepton mass matrix at the electroweak scale in the basis where the charged lepton mass matrix is diagonalized.
For the asymmetries we take the cutoff parameter $\delta = 0.02$.
If $|\lambda_{\tau}| = 10^{-2}$,  $B(\mu^{+} \rightarrow e^{+}\gamma)$
can be $10^{-11}$ and $B(\mu \rightarrow e^{+}e^{+}e^{-})$ can be $10^{-13}$ level, but if $\lambda_{\tau}$ is given by the corresponding KM matrix element, $|\lambda_{\tau}|$ becomes ($3$~--~$5)\times10^{-4}$, so that the branching ratios are smaller by three orders of magnitude.
In Fig.~\ref{fig:su5_1}(c) the ratio of two branching fractions is shown. 
If the photon-penguin contribution dominates over four-fermion ones this ratio is given by Eq. (\ref{eq:ratio}).
We can see that for large parameter region the ratio is enhanced.
In particular, near $m_{\tilde{e}_{R}} = 400$-$600$ GeV almost exact cancellation occurs for the photon-penguin amplitudes \cite{97Hisano}.
In Fig.~\ref{fig:su5_1}(d) $A(\mu^{+}\rightarrow e^{+}\gamma)$ is shown. It is close to 100\% except for small region where the almost exact cancellation occurs.
The P-odd asymmetries $A_{P_{1}}$ and $A_{P_{2}}$ are shown in Fig.~\ref{fig:su5_1}(e) and Fig.~\ref{fig:su5_1}(f).
$A_{P_{1}}$ changes from $-30\%$ to $40\%$ and $A_{P_{2}}$ changes from $-10\%$ to $15\%$.
For $\delta = 0.02$ the asymmetries $A_{P_{1}}$ and $A_{P_{2}}$ are expressed as:
 
\begin{eqnarray}
A_{P_{1}} &\simeq& \frac{3}{2B}\{0.6(C_{1}-C_{2})-0.12(C_{3}-C_{4})+5.6(C_{5}-C_{6}) \nonumber \\
      &      & -4.7(C_{7}-C_{8})+2.5(C_{9}-C_{10})\},
\label{eq:A_P1} \\
A_{P_{2}} &\simeq& \frac{3}{2B}\{0.1(C_{3}-C_{4})+10(C_{5}-C_{6}) \nonumber \\
      &      & +2(C_{7}-C_{8})-1.6(C_{9}-C_{10})\}. 
\label{eq:A_P2}
\end{eqnarray}
In the SU(5) case, because only $g_{3}$, $g_{5}$ and $A_{L}$ have sizable contributions, we obtain the following expressions:

\begin{eqnarray}
A_{P_{1}} &\simeq& \frac{3}{2B}\{0.6|g_{3}|^{2}-0.16|g_{5}|^{2}-5.6|eA_{L}|^{2} \nonumber \\
      &      &+4.7Re(eA_{L}g_{3}^{*})-2.5Re(eA_{L}g_{5}^{*})\}, \label{A1} \\
A_{P_{2}} &\simeq& \frac{3}{2B}\{0.1|g_{5}|^{2}-10|eA_{L}|^{2} \nonumber \\
      &      & -2Re(eA_{L}g_{3}^{*})+1.6Re(eA_{L}g_{5}^{*})\}\label{A2}. 
\end{eqnarray}
In the above formulas we can see that the coefficients for $|A_{L}|^{2}$, $Re(A_{L}g_{3}^{*})$ and $Re(A_{L}g_{5}^{*})$ are large.
Therefore these asymmetries represent the dependence of square of photon-penguin terms and interference terms.
It is interesting to see that we can over-determine the three coupling constants $g_3$,~$g_5$ and $A_L$ from observables $B(\mu^{+}\rightarrow e^{+}\gamma)$, $B(\mu^{+}\rightarrow e^{+}e^{+}e^{-})$, $A_{P_1}$ and $A_{P_2}$ if we assume the SU(5) SUSY GUT without the SUSY CP violating phases.
For example, we can determine $g_3$,~$g_5$ and $A_L$ from the three observables $B(\mu^{+}\rightarrow e^{+}\gamma)$, $B(\mu\rightarrow e^{+}e^{+}e^{-})$ and $A_{P_1}$, then, $A_{P_2}$ can be predicted. 
In addition we should have $A(\mu^{+}\rightarrow e^{+}\gamma)=100\%$ and $A_T=0$.

Next, we include the SUSY CP violating phases and discuss EDM constraints and T-odd asymmetry.
We calculate the electron and neutron EDMs according to Ref. \cite{98Nath}.
Discussion on QCD correction is given in Appendix \ref{sec:nedm}.
For the Hg EDM, we use the result of Ref. \cite{Falk:1999tm}.
$d_{Hg}$ is given:
\begin{eqnarray}
d_{Hg} = -(C^C_{d}-C^C_{u}-0.012C^C_{s})\times3.2\cdot 10^{-2}e,
\end{eqnarray}
where $C^C_{u}$, $C^C_{d}$ and $C^C_{s}$ are chromomagnetic moments discussed in Appendix \ref{sec:nedm}.

\begin{figure}[htbp]
  \begin{center}
    \leavevmode
    \scalebox{.8}{\includegraphics{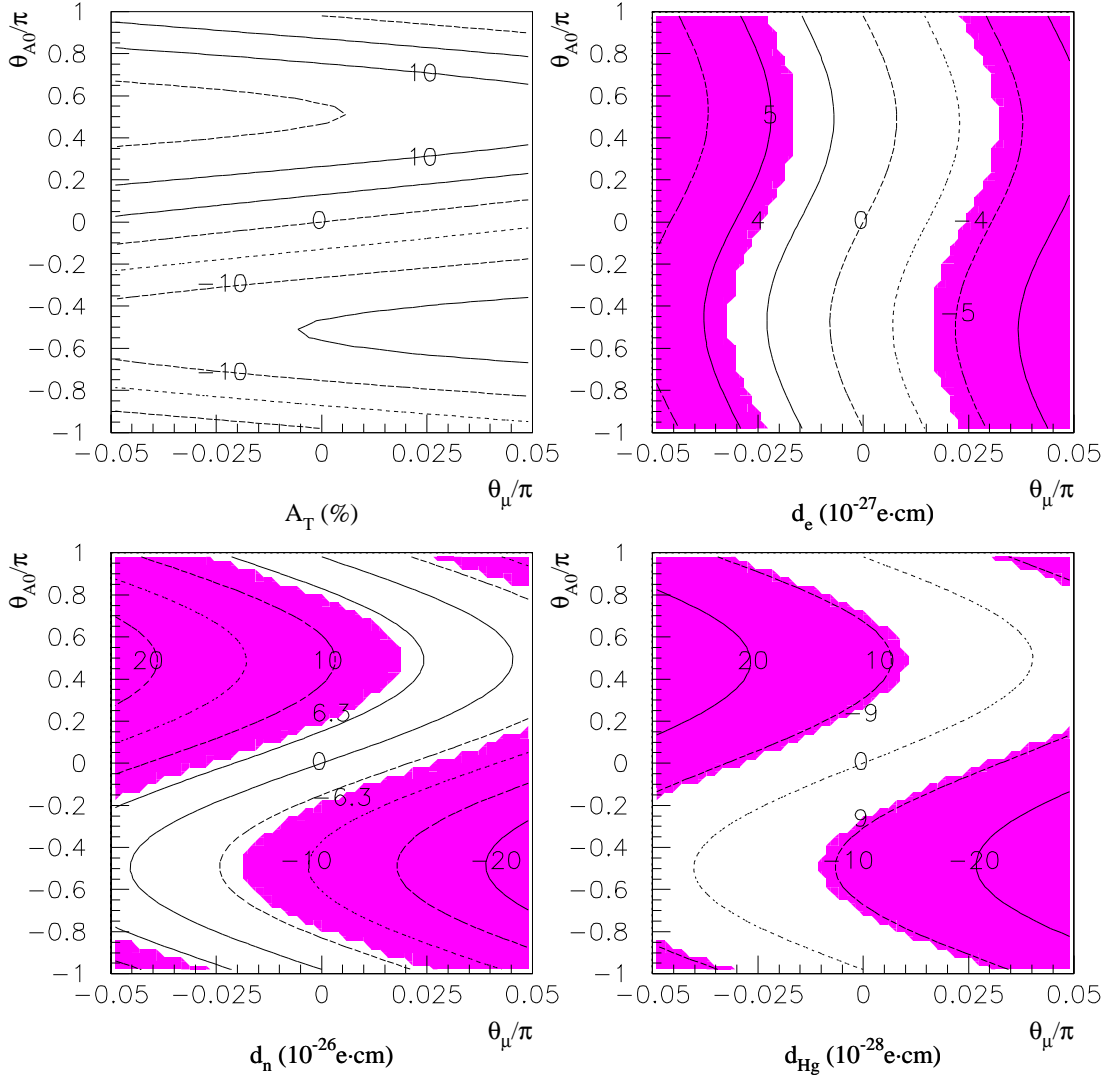}}
    \caption{$\theta_{A_0}$ and $\theta_{\mu}$ dependences on the EDMs and $A_T$.             We take a specific set of SUSY parameters $\tan\beta = 3$, 
              $M_2 = 300$ GeV, 
              $m_{\tilde{e}_R}=650$GeV and $|A_0|=1$ in the parameter region
               $-\pi < \theta_{A_{0}} \leq\pi$ and 
               $-0.05\pi \leq\theta_{\mu}\leq 0.05\pi$.
               Dark shaded regions are excluded by the EDM experiments}
    \label{fig:phase_dep}
  \end{center}
\end{figure}

In order to see $\theta_{A_{0}}$ and $\theta_{\mu}$ dependences on the EDMs and $A_{T}$, we first show these quantities for a specific set of SUSY parameters. In Fig.~\ref{fig:phase_dep}, the electron, neutron and Hg EDMs and $A_{T}$ are shown for $\tan\beta = 3$, $M_{2}=300$GeV, $m_{\tilde{e}_R}=650$GeV, $|A_{0}|=1$ in the parameter region $-\pi < \theta_{A_{0}} \leq\pi$ and $-0.05\pi \leq\theta_{\mu}\leq 0.05\pi$.
The experimental bounds on the EDMs are given by $|d_{e}|<4\times10^{-27}(e\cdot cm)$ \cite{COMMINS94}, $|d_{n}|<0.63\times10^{-25}(e \cdot cm)$ \cite{Harris99} and $|d_{Hg}|<9\times10^{-28}(e \cdot cm)$ \cite{Jacobs93}.
As is well known in Ref. \cite{96FalkOlive} the EDMs are very sensitive to $\theta_{\mu}$, so that $\theta_{\mu}$ is strongly constrained. 
On the other hand $\theta_{A_{0}}$ can be large.
In this particular parameter set, $\theta_{A_{0}}=\frac{\pi}{2}$ is not excluded by three EDM constraints.
Maximum value of the T-odd asymmetry $A_{T}$ in allowed region in this figure is $15\%$.
Note that $A_{T}$ is proportional to $\sin\theta_{A_{0}}$ in a good approximation because the magnitude of $\theta_{\mu}$ is strongly constraint by the EDMs.

\begin{figure}[htbp]
  \begin{center}
    \leavevmode
    \scalebox{.8}{\includegraphics{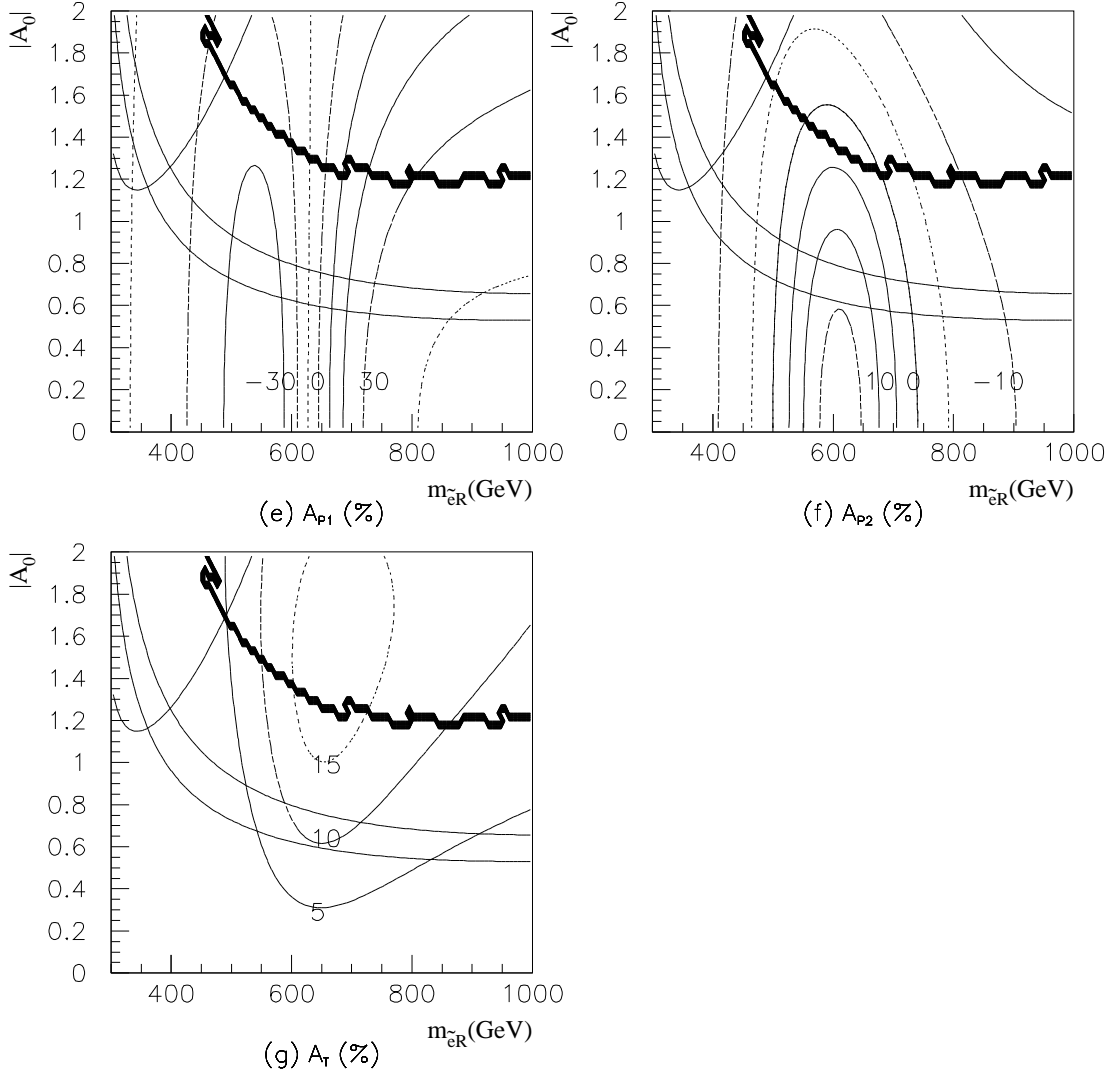}}
    \caption{The obsevables in the SU(5) model with the SUSY CP violating phases 
             in the $m_{\tilde{e}_R}$-$|A_0|$ plane.
             We fix the SUSY parameters as $\tan\beta = 3$,
             $M_2 = 300$ GeV, $\theta_{A_0}=\frac{\pi}{2}$
             and $\theta_{\mu}=0$
              and the top quark mass as 175 GeV.
             (a)-(f) are same as Fig.~\ref{fig:su5_1}.
             (g) The T-odd asymmetry for $\mu^{+} \rightarrow e^{+}e^{+}e^{-}$.
             The cut-off parameter $\delta$ is taken to be 0.02.
             The experimental bounds from the electron, neutron
              and Hg EDMs are also shown in each figure.
             The left upper solid line corresponds to the electron EDM,
             the right upper solid line to the neutron EDM
             and the right lower solid line to the Hg EDM.
             The lower side of each bound is allowed by these experiments. 
             The upper side of the bold line is excluded by the EDM bounds
             even if we allow $\theta_{\mu}$ taking slightly different
              value from $0$.}
    \label{fig:su5_2}
  \end{center}
\end{figure}

\begin{figure}[htbp]
  \begin{center}
    \leavevmode
    \scalebox{.8}{\includegraphics{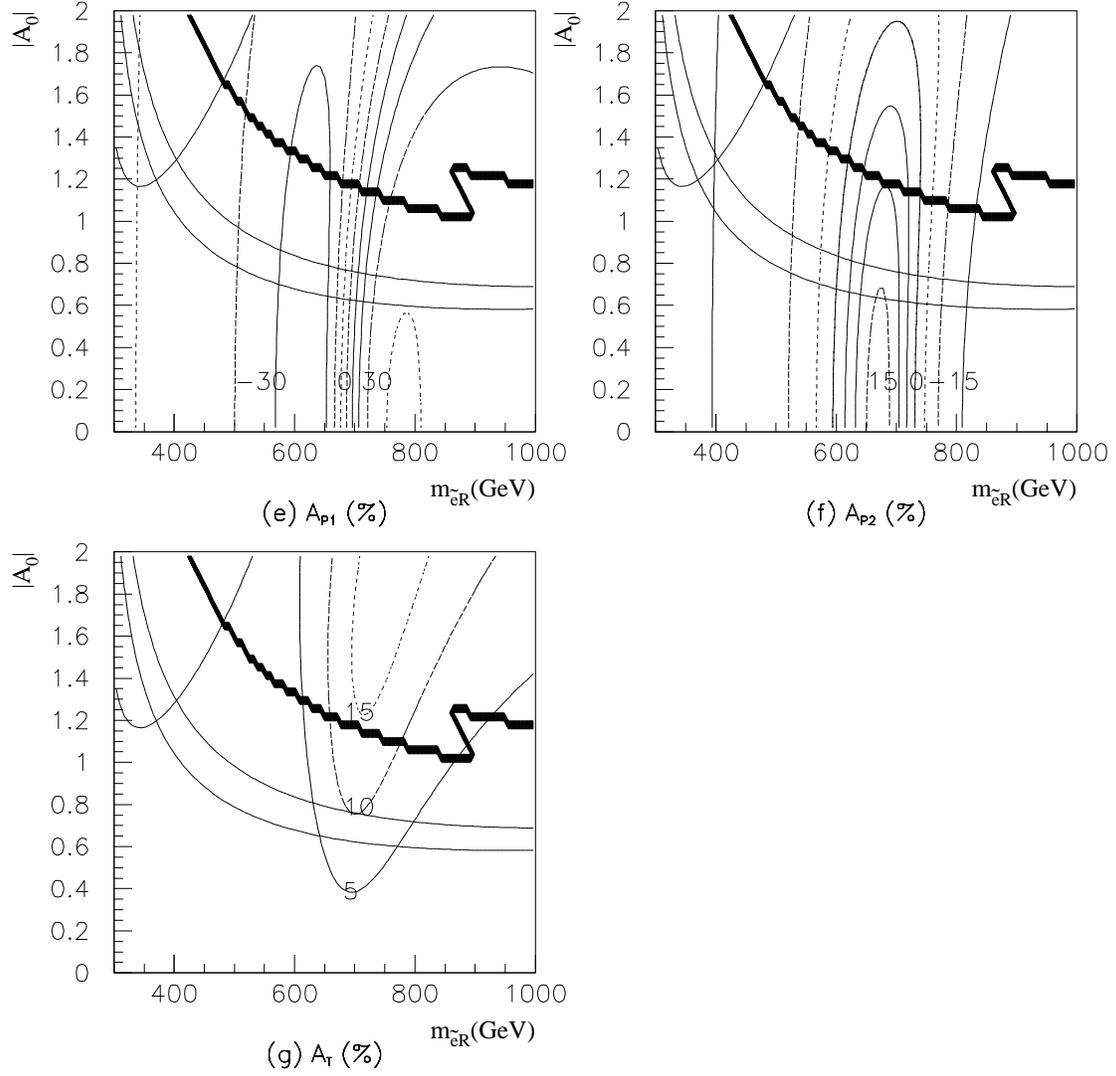}}
    \caption{The obserbables in the SU(5) model for $\tan\beta=10$
             in $m_{\tilde{e}_R}$-$|A_0|$ plane.
             Other parameters are same as in Fig.~\ref{fig:su5_2}.}
    \label{fig:su5_3}
  \end{center}
\end{figure}

In Fig.~\ref{fig:su5_2} we show the quantities in Eq.(\ref{eq:data set}) and $A_{T}$ for $\tan\beta=3$, $M_{2}=300$GeV, $\theta_{A_{0}}=\frac{\pi}{2}$, $\theta_{\mu}=0$. 
We also show the constraints from the electron, neutron and Hg EDMs.
Within the EDM constraints $A_{T}$ can be $10\%$.
As discussed in Fig.~\ref{fig:phase_dep}, when we vary $\theta_{\mu}$ around $\theta_{\mu}=0$, the EDM values change considerably but $A_{T}$ is almost constant.
Therefore the allowed region by the EDM constraints moves in the Fig.~\ref{fig:su5_2} if we take $\theta_{\mu}$ as slightly different value from 0.
On the other hand the contours for branching ratios and the asymmetries in this figure are almost exactly the same. 
In this figure we also show the parameter region which is not allowed by the EDM constraints even if we change $\theta_{\mu}$ around $\theta_{\mu}=0$ for $\theta_{A_{0}}=\frac{\pi}{2}$.
Within the allowed region, the maximum value of $A_{T}$ is $15\%$.
A similar plots are shown for $\tan\beta=10$ in Fig.~\ref{fig:su5_3}.
In this case also the maximum value of $A_{T}$ is about $15\%$.
Note that, in the case with the CP violating phases, we can still determine the complex coupling constants $g_3$, $g_5$ and $A_L$ up to a total phase from the two branching ratios $B(\mu^{+}\rightarrow e^{+}\gamma)$, $B(\mu^{+}\rightarrow e^{+}e^{+}e^{-})$ and three asymmetries $A_{P_1}$, $A_{P_2}$ and $A_T$.

\subsection{SO(10) GUT}
In the SO(10) case, from Eq. (\ref{eq:GUT phases}), there are two physical phases which contribute to the EDMs and $\mu^+ \rightarrow e^+\gamma$ amplitudes.
In the $\mu^+ \rightarrow e^+\gamma$ amplitudes the term proportional to $m_{\tau}$ has a dependence of $e^{i(\phi_3-\phi_2)}(V_{KM}^0)_{32}\{(V_{KM}^0)_{33}^*\}^2(V_{KM}^0)_{31}$ and other contributions are proportional to $(V_{KM}^0)_{32}^*(V_{KM}^0)_{31}$.
Therefore, the branching ratio $\mu^+ \rightarrow e^+\gamma$ depends on the relative phase of two terms.
In the followings we consider the case where there is no relative phase so that the amplitude is proportional to $\lambda_{\tau}$.
Also we do not consider EDM constraints from Eq. (\ref{eq:EDM relation}) explicitly since this can be suppressed when $\phi$ is small.

\begin{figure}[htbp]
  \begin{center}
    \leavevmode
    \scalebox{.8}{\includegraphics{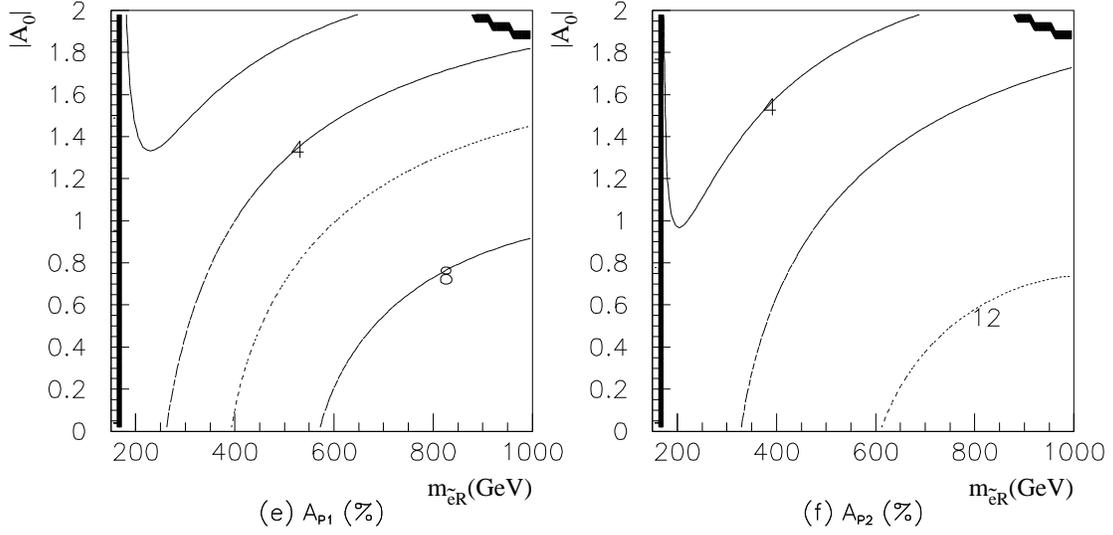}}
    \caption{The observables in the SO(10) model without the SUSY CP
              violating phase in $m_{\tilde{e}_R}$-$|A_0|$ plane.
              The input parameters are same as in Fig.~\ref{fig:su5_1}. 
              The upper right corner bounded by the bold line is excluded by
               phenomenological constraints and the small $m_{\tilde{e}_R}$
                region bounded by the left bold line is not allowed in the
              minimal SUGRA model.}
    \label{fig:so10_1}
  \end{center}
\end{figure}

In Fig.~\ref{fig:so10_1} the branching ratios and the asymmetries are shown for the SO(10) model.
We first show the case without the SUSY CP violating phases. Input SUSY parameters are taken as $\tan\beta=3$, $M_{2}=150$GeV, $\theta_{A_{0}}=0$ and $\theta_{\mu}=0$.
We see that $B(\mu^{+} \rightarrow e^{+} \gamma)/|\lambda_{\tau}|^{2}$ can be $10^{-3}$.
This value is enhanced by 2-4 orders of magnitude compared to the SU(5) case.
 The ratio of two branching fractions is almost constant because the photon-penguin diagrams give dominant contributions to $\mu^{+} \rightarrow e^{+}e^{+}e^{-}$. 
The $\mu^{+}\rightarrow e^{+}\gamma$ asymmetry $A(\mu^{+}\rightarrow e^{+}\gamma)$ varies from $-20\%$ to $-90\%$.
This is in contrast to the previous belief that $A_L$ and $A_R$ have a similar magnitude in this model.
Although the diagram proportional to $m_{\tau}$ gives the same contribution to the $A_{L}$ and $A_{R}$, there is a chargino loop diagram which only contributes to $A_{R}$.
In spite of no $m_{\tau}$ enhancement, the contribution from the latter diagram can be comparable to that from the former one, especially when the slepton mass is larger than the chargino mass.
In Fig.~\ref{fig:so10_1}(e),~(f) the P-odd asymmetries for $\mu^{+}\rightarrow e^{+}e^{+}e^{-}$ are shown and these asymmetries are small compared to the SU(5) case.
$A_{P1}$ is less than $10\%$ and $A_{P2}$ is less than $14\%$.
In this case $C_5$ and $C_6$ terms dominate in Eqs. (\ref{eq:A_P1}) and (\ref{eq:A_P2}) so that these asymmetries are proportional to $A(\mu^+\rightarrow e^+\gamma)$.
We have also investigated the case with $\tan\beta=10$.
We found the parity asymmetry for $\mu^+ \rightarrow e^+\gamma$ and $A_{P1}$, $A_{P2}$ have a similar magnitudes as Fig.~\ref{fig:so10_1}, namely, $A(\mu^+ \rightarrow e^+\gamma)$ varies $-20\%$~--~$-100\%$, $A_{P1}$ varies $2\%$~--~$10\%$ and $A_{P2}$ varies $4\%$~--~$16\%$ in the same parameter space. 

\begin{figure}[htbp]
  \begin{center}
    \leavevmode
    \scalebox{.8}{\includegraphics{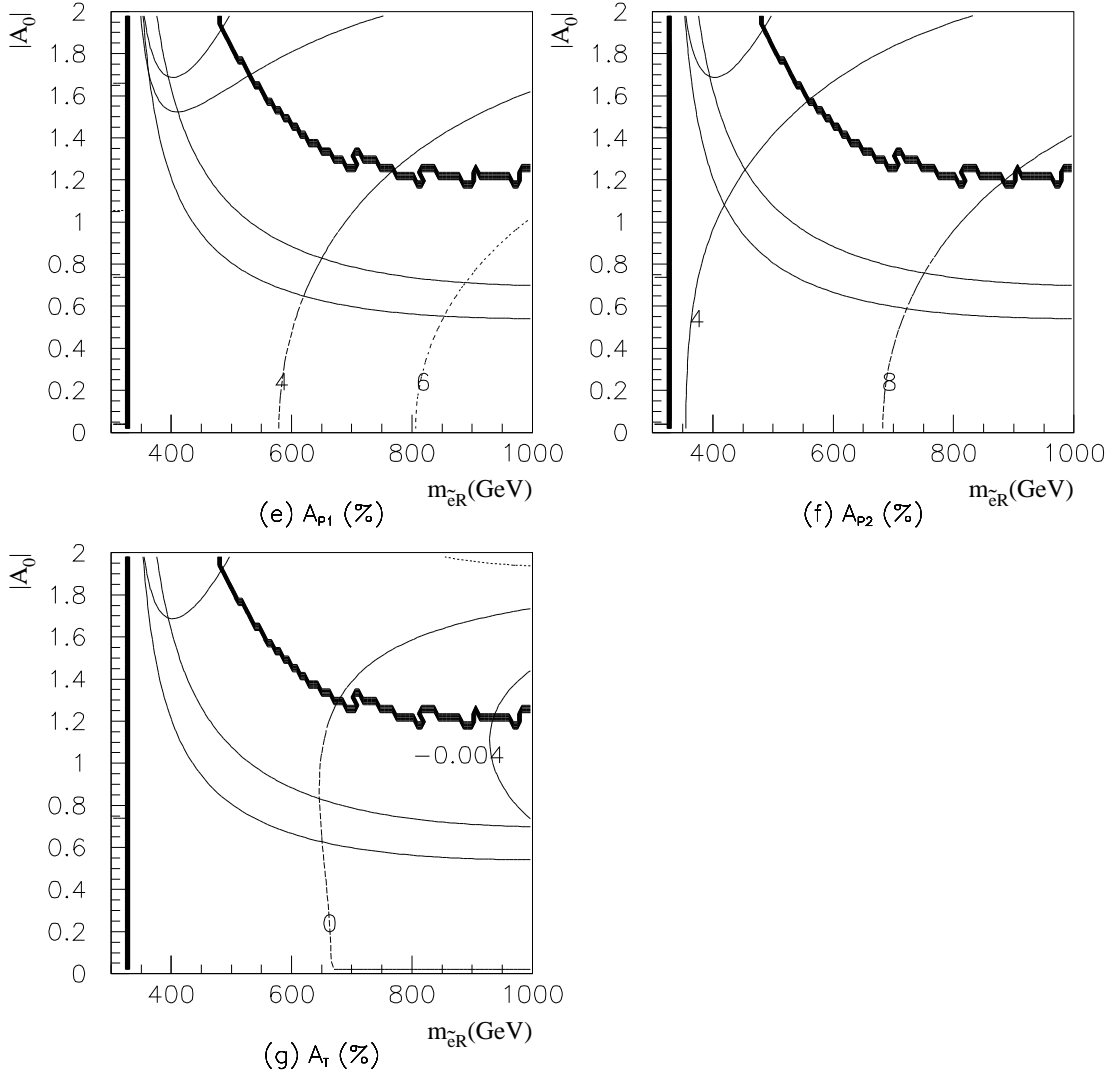}}
    \caption{The obsevables in the SO(10) model
              with the SUSY CP violating phase
              in $m_{\tilde{e}_R}$-$|A_0|$ plane. 
              The input parameters are same as 
              in Fig.~\ref{fig:su5_2}.
              The small $m_{\tilde{e}_R}$ region bounded by the left
               bold line is not allowed in the minimal SUGRA model.
              The upper right bold line shows the same bound as in
               Fig.~\ref{fig:su5_2}.}
    \label{fig:so10_2}
  \end{center}
\end{figure}

In Fig.~\ref{fig:so10_2} we consider the case with the SUSY CP violating phase and take input parameters as $\tan\beta=3$, $M_{2}=300$GeV, $\theta_{A_{0}}=\frac{\pi}{2}$ and $\theta_{\mu}=0$.
The branching ratio and other asymmetries have similar magnitudes compared to the case in Fig.~\ref{fig:so10_1}.
We can see that the T-odd asymmetry $A_{T}$ is less than $0.01\%$ because only the photon-penguin amplitude becomes large.

Some remarks are in order:
\begin{enumerate}
\item When we take into account the phase in Eq. (\ref{eq:relative phase}), the EDM is generated as discussed in Eq. (\ref{eq:EDM relation}).
We note that the T-odd asymmetry cannot be large even in such a case because the photon-penguin diagram dominates over the four-fermion contributions.
\item If the $\mu^+ \rightarrow e^+\gamma$ asymmetry is sizable, the simple relationship between the EDM and the $\mu^+ \rightarrow e^+\gamma$ branching ratio as in Eq. (\ref{eq:EDM relation}) does not hold. 
This is because the EDM amplitude is no longer proportional to the $\mu^+ \rightarrow e^+\gamma$ amplitude due to the chargino loop contribution.
\item Even if we include relative phases between the term proportional to $m_{\tau}$ and other contribution in the $\mu^+ \rightarrow e^+ \gamma$ amplitude, we expect large $A(\mu^+ \rightarrow e^+\gamma)$ as long as two contributions have a similar magnitude. We have numerically checked that the asymmetry varies $0\%$~--~$-100\%$ by changing the relative phase.
\end{enumerate}

\subsection{Differential branching ratio and asymmetries}
\begin{figure}[htbp]
  \begin{center}
    \leavevmode
    \scalebox{.8}{\includegraphics{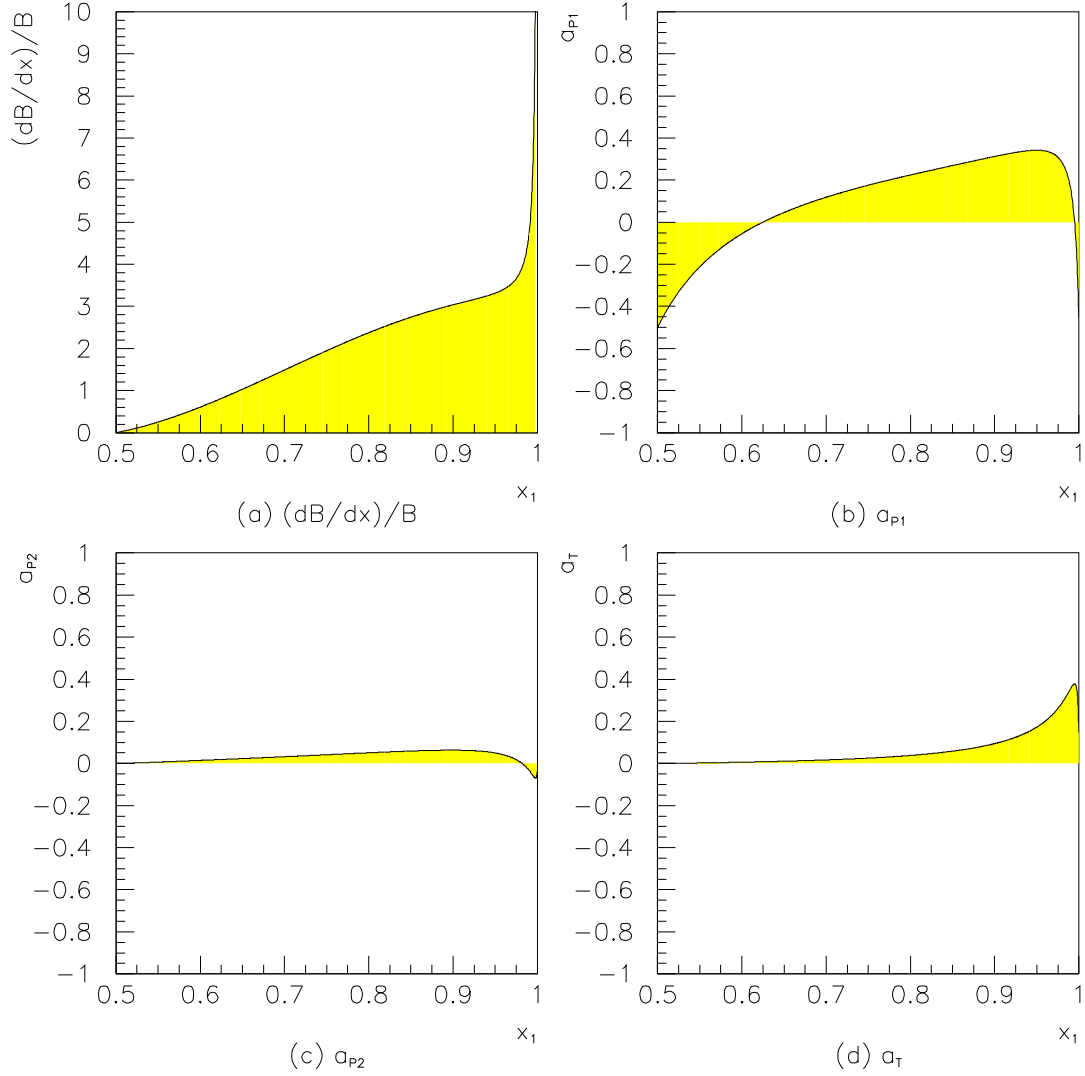}}
    \caption{The differential branching ratio and asymmetries for 
             the $\mu^{+}\rightarrow e^{+}e^{+}e^{-}$ process in the SU(5) model as a
              function of $x_1$ which
             is a larger energy of decay positrons ($\frac{2E_1}{m_{\mu}}$). 
             We fix the SUSY parameters as  $\tan\beta = 3$,  
             $M_2 = 300$ GeV, $m_{\tilde{e}_R}=700$GeV,
             $|A_0| = 0.5$, $\theta_{A_0}=\frac{\pi}{2}$
              and $\theta_{\mu}=0$.
             (a) The differential branching ratio for
              the $\mu^{+}\rightarrow e^{+}e^{+}e^{-}$ 
             normalized by the total branching ratio.
             (b) The differential P-odd asymmetry $a_{P_1}$.
             (c) The differential P-odd asymmetry $a_{P_2}$.
             (d) The differential T-odd asymmetry $a_T$.}
    \label{fig:su5_dalitz}
  \end{center}
\end{figure}

\begin{figure}[htbp]
  \begin{center}
    \leavevmode
    \scalebox{.8}{\includegraphics{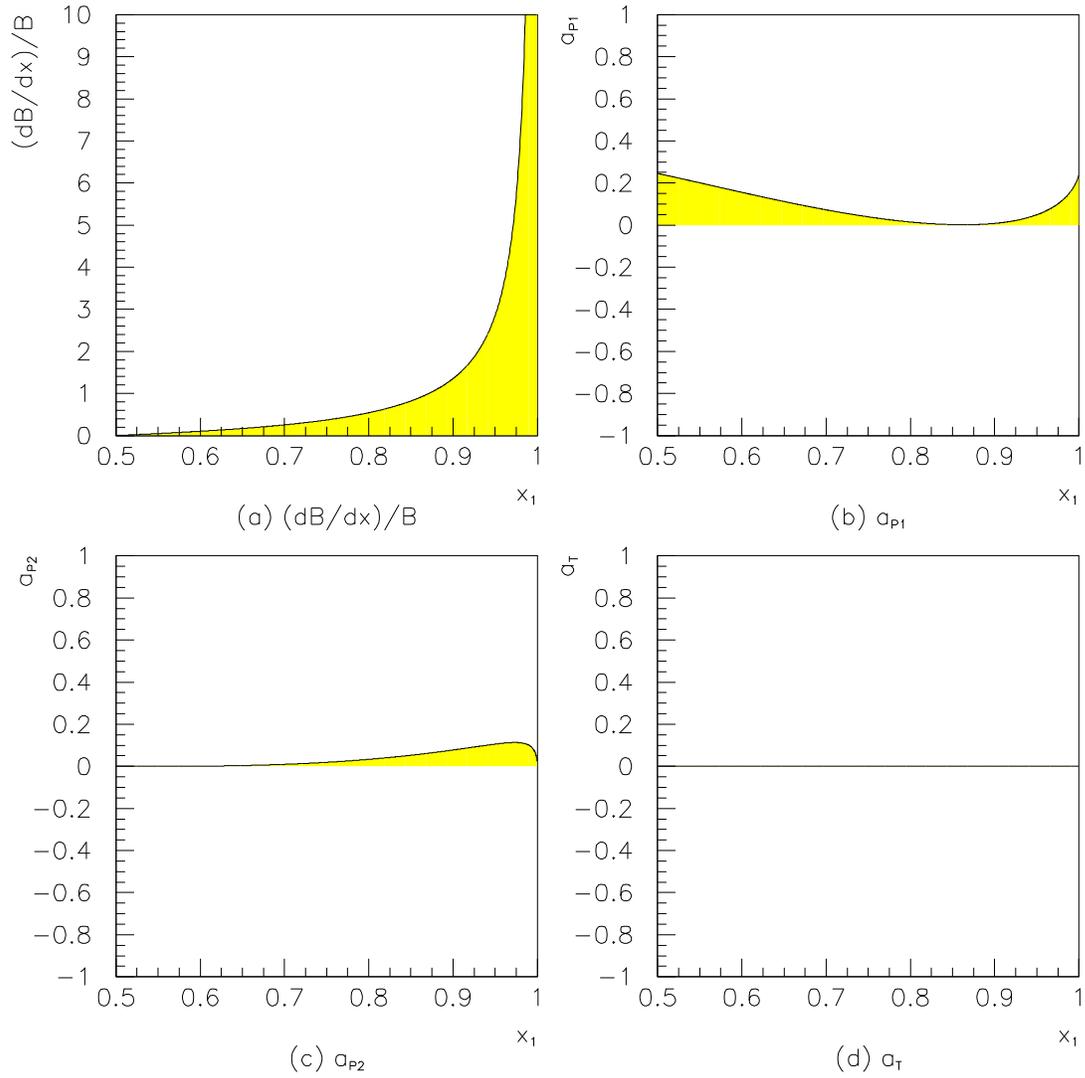}}
    \caption{The differential branching ratio and asymmetries for 
             the $\mu^{+}\rightarrow e^{+}e^{+}e^{-}$ process in the SO(10)
              model as a function of $x_1$. The input parameters are
              same as in Fig.~\ref{fig:su5_dalitz}.}
    \label{fig:so10_dalitz}
  \end{center}
\end{figure}

Up to now we only discussed the integrated branching ratio and asymmetries of $\mu^+ \rightarrow e^+e^+e^-$.
In the actual experiment, the differential quantities are useful to distinguish different models.
For example, in Fig.~\ref{fig:su5_dalitz} and \ref{fig:so10_dalitz}, we show the differential branching ratio and asymmetries for a particular parameter set in the SU(5) and SO(10) models. 
$\frac{dB}{dx_1}$, $a_{P_{1}}$, $a_{P_{2}}$ and $a_{T}$  are plotted for the parameter set of $\tan \beta = 3$, $m_{\tilde{e}_{R}}=700$ GeV, $M_{2}=300$ GeV, $|A_{0}|=0.5$, $\theta_{A_{0}}=\frac{\pi}{2}$ and $\theta_{\mu}=0$.
We can see clear difference between the SU(5) and SO(10) models.
The differential branching has the steep peak near $x_1=1$ for the SO(10) case whereas the distribution is broader for the SU(5) case.
This is because the photon-penguin contribution has $\frac{1}{1-x_1}$ behavior near $x_1=1$ and the four-fermion operators give a broad spectrum.
 We also see the T-odd asymmetry has the peak at $x_1$ close to 1. 
This fact arises from the $\frac{1}{\sqrt{1-x_1}}$ behavior in the $\gamma_3$ and $\gamma_4$ near $x_1=1$.
Because of this feature of distribution, we have chosen $\delta = 0.02$ to optimize the T-odd asymmetry. 

\section{Conclusion} 
 We developed the model-independent formalism for the process $\mu^{+} \rightarrow e^{+} \gamma$ and $\mu^{+} \rightarrow e^{+}e^{+}e^{-}$ with polarized muon and defined convenient observables such as the P-odd and T-odd asymmetries.
Using explicit calculation based on the SU(5) and SO(10) SUSY GUT, we show that various combination of LFV coupling constants can be determined from the measurement of branching ratio and asymmetries.
 In the SO(10) case the P-odd asymmetry in $\mu^{+}\rightarrow e^{+}\gamma$ varies from $-20\%$ to $-100\%$ whereas it is $+100\%$ for the SU(5) case.
With the branching ratio and the P-odd asymmetries in the $\mu^{+}\rightarrow e^{+}e^{+}e^{-}$ process, we can over-determine the coupling constants in the effective Lagrangian in the SU(5) SUSY GUT if there is no SUSY CP violating phases. 
 We also calculated the T-odd asymmetry in the $\mu^{+}\rightarrow e^{+}e^{+}e^{-}$ process with the SUSY CP violating phases and compare it with the neutron, electron and Hg EDMs. 
In the SU(5) case 
we can still determine these coupling constants using additional information of the T-odd asymmetry. The T-odd asymmetry can reach 15\% within the constraints of the EDMs.
In the SO(10) case
the T-odd asymmetry is small as a result of the dominance of photon-penguin diagram.
These results are summarized in Table \ref{tbl:results}.
We stress that although the magnitude of the branching ratio has a large uncertainty due to the unknown parameter $\lambda_{\tau}$, asymmetries and the ratio of two branching ratios are independent of this ambiguity.
Thus these quantities are useful to distinguish different models.

The experimental prospects for measuring these quantities depend on the branching ratio.
For the SO(10) model we expect the $\mu^{+}\rightarrow e^{+}\gamma$ branching ratio can be $10^{-12}$ when the $\lambda_{\tau}$ is given by the corresponding KM matrix elements.
In such a case the $\mu^{+}\rightarrow e^{+}\gamma$ asymmetry can be measurable in an experiment with a sensitivity of $10^{-14}$ level.
For the SU(5) model, to get the $\mu^{+}\rightarrow e^{+}\gamma$ branching ratio of order $10^{-12}$ and $\mu^{+}\rightarrow e^{+}e^{+}e^{-}$ branching ratio of $10^{-14}$, we have to assume $\lambda_{\tau}$ is larger than a several times $10^{-3}$.
If the branching ratio turns out to be as large, the $\mu^{+}\rightarrow e^{+}e^{+}e^{-}$ experiments with a sensitivity of $10^{-16}$ level could reveal various asymmetries.
Because various asymmetries are defined with respect to muon polarization, experimental searches for $\mu^{+}\rightarrow e^{+}\gamma$ and $\mu^{+} \rightarrow e^{+}e^{+}e^{-}$ with polarized muons are very important to uncover the nature of the LFV interactions.

The authors would like to thank Y. Kuno and J. Hisano for useful discussion.
The work of Y.O. was supported in part by the Grant-in-Aid of the Ministry of Education, Science, Sports and Culture, Government of Japan (No.09640381),Priority area ``Supersymmetry and Unified Theory of Elementary Particles'' (No. 707), and ``Physics of CP Violation'' (No.09246105).

\begin{table}[h]
\begin{center}
\begin{tabular}{|c|c|c|} \hline
 & SU(5) SUSY GUT & SO(10) SUSY GUT \\ \hline
\hline
$A(\mu^{+}\rightarrow e^{+}\gamma)$ & $+100\%$ & $-20\%$~--~$-100\%$ \\ \hline
$\frac{B(\mu^{+}\rightarrow e^{+}e^{+}e^{-})}{B(\mu^{+}\rightarrow e^{+}\gamma)}$
 & $0.007$~--~$O(1)$
 & constant~($\sim 0.0062$) \\ \hline
$A_{P_1}$
 & $-30\%$~--~$+40\%$
 & $\lsim 10\%$ \\ \hline
$A_{P_2}$
 & $-20\%$~--~$+20\%$
 & $\lsim 15\%$ \\ \hline
$|A_T|$
 & $\lsim 15\%$
 & $\lsim 0.01\%$ \\ \hline
\end{tabular}
\caption{Summary of the results\label{tbl:results}}
\end{center}
\end{table}

\vspace{2em}

\newpage

\appendix
\section{Branching Ratio and Asymmetries}
\label{sec:asym}

 In this appendix, we give kinematic functions which appear in the calculation of branching ratio and asymmetries. 

\begin{eqnarray}
F_{i}(x) & \equiv & 2\int_{1-x}^{x} dx_{2} \alpha_{i}(x,x_{2}), 
                    \\
G_{i}(x) & \equiv & 2\int_{1-x}^{x} dx_{2} \beta_{i}(x,x_{2}),
                    \\ 
H_{i}(x) & \equiv & -2\int_{1-x}^{x} dx_{2} \gamma_{i}(x,x_{2}), 
\end{eqnarray}

\begin{eqnarray}
F_{1}(x) & = & -\frac{8}{3}(4x-5)(2x-1)^2, 
 \\ 
F_{2}(x) & = & -\frac{2}{3}(2x-1)(8x^2-8x-1),
 \\ 
F_{3}(x) & = & 16\ln(\frac{x}{1-x})(2x^{2}-2x+1)
                         +\frac{32}{3}\frac{(2x-1)(x^{2}-x+1)}{1-x},
 \\ 
F_{4}(x) & = & 32(2x-1)^{2},
 \\ 
F_{5}(x) & = & -8(2x-1)(2x-3),
 \\
G_{1}(x) & = & -16\,(1-x)^{2}\ln 2(1-x) 
                        -\frac{2}{3}\,(2x-1)(8x^{2}-32x+23),
 \\
G_{2}(x) & = & -16(2x^2-2x-7)\ln 2(1-x)
               +16(2x^2-2x+1)\ln 2x \nonumber\\
         &   & +\frac{32}{3}\frac{(2x-1)(x^2-13x+13)}{1-x},
 \\
H_{1}(x) & = & 2(6-5x)(1-x)\sqrt{2x-1} \nonumber \\
         &   &           -(7x^{2}-24x+16)\sqrt{1-x}
                          \arccos(\frac{2-3x}{x}) \nonumber \\
         &   &           +16(1-x)^{2}
                          \arccos(\frac{1-x}{x})
                       \},
 \\
H_{2}(x) & = &  -16(6-x)\sqrt{2x-1} \nonumber \\
         &   &  -8\frac{5x^{2}+8x-16}{\sqrt{1-x}}\arccos(\frac{2-3x}{x})
                -128\arccos(\frac{1-x}{x}),
 \\
H_{3}(x) & = &  -\frac{4}{3}\sqrt{2x-1}(17x^{2}-24x+4)
                +2\frac{(7x-6)x^{2}}{\sqrt{1-x}}
                          \arccos(\frac{2-3x}{x}),
 \\
H_{4}(x) & = & +\frac{2}{3}\sqrt{2x-1}(17x^{2}-30x+16) \nonumber \\
         &   & -\frac{(7x^{2}-16x+8)x}{\sqrt{1-x}}
                          \arccos(\frac{2-3x}{x}).
\end{eqnarray}

\begin{eqnarray}
I_{i}[\delta] & = & \int_{\frac{1}{2}}^{1-\delta}
                     dx F_{i}(x), \\
J_{i}[\delta] & = & \int_{\frac{1}{2}}^{1-\delta}
                     dx G_{i}(x){dx}, \\
K_{i}[\delta] & = & \int_{\frac{1}{2}}^{1-\delta}
                     dx H_{i}(x){dx},
\end{eqnarray}

\begin{eqnarray}
I_{1}[\delta] & = & \frac{2}{3}(1+2\delta)(1-2\delta)^3,
 \\
I_{2}[\delta] & = & \frac{1}{3}(1+2\delta-2\delta^2)(1-2\delta)^2,
 \\
I_{3}[\delta] & = & \frac{16}{3}(1-\delta)(2-\delta+2\delta^2)
                    \ln(\frac{1-\delta}{\delta})
                    -\frac{8}{9}(1-2\delta)(13-4\delta+4\delta^2),
 \\
I_{4}[\delta] & = & \frac{16}{3}(1-2\delta)^3,
 \\
I_{5}[\delta] & = & \frac{8}{3}(1+\delta)(1-2\delta)^2,
 \\
J_{1}[\delta] & = & -\frac{1}{9}-\frac{2}{3}\delta+6\delta^2
                    +\frac{16}{3}(\ln 2\delta -\frac{4}{3})\delta^3
                    -\frac{8}{3}\delta^4,
 \\
J_{2}[\delta] & = & -\frac{16}{3}(2+21\delta+3\delta^2-2\delta^3)\ln 2\delta
                   +\frac{16}{3}(1-\delta)(2-\delta+2\delta^2)\ln 2(1-\delta)
                    \nonumber \\
              &   &-\frac{8}{9}(1-2\delta)(49+68\delta+4\delta^2),
 \\
K_{1}[\delta] & = & \frac{4}{315}(8+8\delta-93\delta^{2}-225\delta^{3})
                    \sqrt{1-2\delta} \nonumber \\
              &   & -\frac{2}{3}\delta^{\frac{3}{2}}(1-6\delta-3\delta^{2})
                    \arccos(\frac{3\delta-1}{1-\delta}) \nonumber \\
              &   & -\frac{16}{3}\delta^{3}\arccos(\frac{\delta}{1-\delta}),
 \\
K_{2}[\delta] & = & \frac{32}{5}(4+9\delta+\delta^{2})\sqrt{1-2\delta}
                    -16\sqrt{\delta}(3+6\delta-\delta^{2})
                     \arccos(\frac{3\delta-1}{1-\delta}) \nonumber \\
              &   & +128\delta\arccos(\frac{\delta}{1-\delta}),
 \\
K_{3}[\delta] & = & \frac{8}{105}\sqrt{1-2\delta}
                    (48-57\delta-68\delta^{2}+85\delta^{3}) \nonumber \\
              &   & -4(1-\delta)^{3}\sqrt{\delta}
                     \arccos(\frac{3\delta-1}{1-\delta}),
 \\
K_{4}[\delta] & = & \frac{4}{105}\sqrt{1-2\delta}
                    (64-41\delta+26\delta^{2}-85\delta^{3}) \nonumber \\
              &   & -2(1-\delta-\delta^{2}+\delta^{3})\sqrt{\delta}
                    \arccos(\frac{3\delta-1}{1-\delta}).
\end{eqnarray}

\section{LFV effective coupling constants in MSSM}
\subsection{MSSM Lagrangian}

 We first fix our notations of the MSSM for the numerical calculation. 
Using
\begin{equation}
v \equiv \sqrt{2({\langle H^{0}_{2} \rangle}^{2}
                +{\langle H^{0}_{1} \rangle}^{2})}, 
\end{equation}
and
\begin{equation}
\tan\beta \equiv \frac{\langle H^{0}_{2} \rangle}{\langle H^{0}_{1} \rangle},
\end{equation}
the charged lepton mass matrix is given:
\begin{equation}
(m_{e})_{ij} = -(y_{e})_{ij}\frac{v}{\sqrt{2}}\cos\beta.
\end{equation}
 The neutralino and chargino mass matrices are written as follows:

\begin{eqnarray}
&& {\cal L}_{N}  = -\frac{1}{2}
\left(
\begin{array}{llll}
\overline{\widetilde{B}}_{R} & \overline{\widetilde{W}_{3}}_{R} &
 \overline{\widetilde{H}^{0}_{1}}_{R}^{c} & 
\overline{\widetilde{H}^{0}_{2}}_{R}^{c} 
\end{array} 
\right)
{\cal M}_{N}
\left(
\begin{array}{l}
\widetilde{B}_{L} \\
\widetilde{W}_{3L}\\
\widetilde{H}_{1L}^{0} \\ 
\widetilde{H}_{2L}^{0}
\end{array}
\right) + h.c. ,
 \nonumber \\
&& {\cal M}_{N}  = 
 \left(
\begin{array}{cccc}
M_{1} & 0     & -m_{z}\sin\theta_{W}\cos\beta
      & m_{z}\sin\theta_{W}\sin\beta \\
0     & M_{2} & m_{z}\cos\theta_{W}\cos\beta
      & -m_{z}\cos\theta_{W}\sin\beta \\
        -m_{z}\sin\theta_{W}\cos\beta & m_{z}\cos\theta_{W}\cos\beta
      & 0 & -\mu \\
        m_{z}\sin\theta_{W}\sin\beta & -m_{z}\cos\theta_{W}\sin\beta
      & -\mu & 0
\end{array}
\right),\nonumber \\
\end{eqnarray}

\begin{eqnarray}
&& {\cal L}_{C} = 
-\left(
\begin{array}{ll}
\overline{\widetilde{W}^{-}}_{R} &
\overline{\widetilde{H}^{-}_{2}}_{R}^{c}  
\end{array} 
\right)
{\cal M}_{C}
\left(
\begin{array}{l}
\widetilde{W}^{-}_{L}\\
\widetilde{H}_{1L}^{-}
\end{array}
\right)  + h.c.,
 \nonumber \\
&& {\cal M}_{C} = 
\left(
\begin{array}{cc}
M_{2} & \sqrt{2}m_{W}\cos\beta \\
\sqrt{2}m_{W}\sin\beta & \mu 
\end{array}
\right) . 
\end{eqnarray}
They are diagonalized with unitary matrices as follows:

\begin{equation}
O_{N} {\cal M}_{N} O_{N}^{T} = diag(m_{\widetilde{\chi}^{0}_{1}},
                                m_{\widetilde{\chi}^{0}_{2}},
                                m_{\widetilde{\chi}^{0}_{3}},
                                m_{\widetilde{\chi}^{0}_{4}}),
\end{equation}
\begin{eqnarray}
O_{NR}  &\equiv&  O_{N},
 \\
O_{NL}  &\equiv&  {O_{N}}^{*},
\end{eqnarray}
\begin{equation}
O_{CR} {\cal M}_{C} O_{CL}^{\dag} = diag(m_{\tilde{\chi}^{-}_{1}},
                                m_{\tilde{\chi}^{-}_{2}}).
\end{equation}
The slepton mass matrices are written  as follows:

\begin{eqnarray}
&& {\cal L}_{\tilde{e}} =
-\left(
\begin{array}{ll}
\widetilde{L}^{-\dag} & \widetilde{E}^{\dag}
\end{array}
\right)
m_{\tilde{e}}^{2}
\left(
\begin{array}{l}
\widetilde{L}^{-} \\
\widetilde{E}
\end{array}
\right),
 \nonumber \\
&& m_{\tilde{e}}^{2} =
\left(
\begin{array}{cc}
m^{2}_{L} + m_{e}^{\dag}m_{e}
+m_{z}^{2}\cos 2\beta(-\frac{1}{2}+\sin^{2}\theta_{W}) & 
\frac{v}{\sqrt{2}}\cos\beta(m_0 A_e+y_e \mu^* \tan\beta)^{\dag} \\
\frac{v}{\sqrt{2}}\cos\beta(m_0 A_e+y_e \mu^* \tan\beta) & 
m^{2}_{E}+m_{e}m_{e}^{\dag}
-m_{z}^{2}\cos 2\beta\sin^{2}\theta_{W}
\end{array}
\right), \nonumber \\
\end{eqnarray}

\begin{eqnarray}
&& {\cal L}_{\tilde{\nu}} =
-\tilde{L}^{0\dag}
m_{\tilde{\nu}}^{2}
\tilde{L}^{0},
\nonumber  \\
&& m_{\tilde{\nu}}^{2} = m^{2}_{L} -\frac{1}{2}m_{z}^{2}\cos 2\beta. 
\end{eqnarray}
They are diagonalized with unitary matrices as follows:

\begin{eqnarray}
U_{e}m_{\tilde{e}}^{2}U_{e}^{\dag} &=& diag(m_{\tilde{e}_{1}}^{2},
                                          m_{\tilde{e}_{2}}^{2},
                                          m_{\tilde{e}_{3}}^{2},
                                          m_{\tilde{e}_{4}}^{2},
                                          m_{\tilde{e}_{5}}^{2},
                                          m_{\tilde{e}_{6}}^{2}),
 \\
U_{\nu}m_{\tilde{\nu}}^{2}U_{\nu}^{\dag} &=& diag(m_{\tilde{\nu}_{1}}^{2},
                                          m_{\tilde{\nu}_{2}}^{2},
                                          m_{\tilde{\nu}_{3}}^{2}).
\end{eqnarray}
The neutralino and chargino vertices for leptons and sleptons are written as follows:

\begin{eqnarray}
{\cal L}  & \equiv & \overline{e_{i}}(N^{L}_{iAX}P_{L}+N^{R}_{iAX}P_{R})
                       \tilde{\chi}^{0}_{A}\tilde{e}_{X} \nonumber \\
          &   &  + \overline{e_{i}}(C^{L}_{iAX}P_{L}+C^{R}_{iAX}P_{R})
                       \tilde{\chi}^{-}_{A}\tilde{\nu}_{X} \nonumber \\ 
              &   & +h.c.,
\end{eqnarray}

\begin{eqnarray}
N^{L}_{iAX} & = & -g\{ \sqrt{2}\tan\theta_{W}
                       (O_{NL})^{*}_{A1}(U_{e})_{X i+3}^{*}
                      +\frac{(m_{e})_{ij}}{\sqrt{2}m_{W}\cos\beta}
                       (O_{NL})^{*}_{A3}(U_{e})_{Xj}^{*} \},
 \\
N^{R}_{iAX} & = & -g[-\frac{1}{\sqrt{2}}
                       \{(O_{NR})_{A2}^{*}
                   +\tan\theta_{W}(O_{NR})_{A1}^{*} \}
                       (U_{e})_{Xi}^{*} \nonumber \\
            &   &  +\frac{(m_{e}^{\dag})_{ij}}{\sqrt{2}m_{W}\cos\beta}
                       (O_{NR})_{A3}^{*}(U_{e})_{Xj+3}^{*} ]. 
\end{eqnarray}

\begin{eqnarray}
C^{L}_{iAX} & = & g\frac{(m_{e})_{ij}}{\sqrt{2}m_{W}\cos\beta}
                  (O_{CL})_{A2}^{*}(U_{\nu})_{Xj}^{*},
 \\
C^{R}_{iAX} & = & -g(O_{CR})_{A1}^{*}(U_{\nu})_{Xi}^{*}, 
\end{eqnarray}

\subsection{LFV effective coupling constants}
\label{sec:couplings}

 The formulas of effective coupling constants for $\mu \rightarrow e\gamma$ and $\mu \rightarrow 3e$ processes written in the MSSM variables are given in Ref.\cite{96Hisano}. 
 We present these formulas for completeness with taking care of the CP violating phases. 

 Each coupling constant is divided into a neutralino-charged-slepton-loop contribution and chargino-sneutrino-loop contribution.
The four-fermi coupling constants are given as follows:

\begin{equation}
g_{i} = g_{i}^{n} + g_{i}^{c}\;(i=1-6).
\end{equation}
The coupling constant $g_{1}$ comes only from box diagrams.

\begin{eqnarray}
g_{1}^{n} & = & -\frac{\sqrt{2}}{64\pi^2 G_F}\sum_{A,B=1}^{4}\sum_{X,Y=1}^{6}
        (N^{L}_{2AX} N^{R*}_{1AY} N^{L}_{1BY} N^{R*}_{1BX}
         -2N^{L}_{2AX} N^{L}_{1AY} N^{R*}_{1BY} N^{R*}_{1BX})
          \nonumber \\
          &   &m_{\tilde{\chi}^{0}_{A}}m_{\tilde{\chi}^{0}_{B}}
          d_{0}({m_{\tilde{\chi}^{0}_{A}}}^{2},
                {m_{\tilde{\chi}^{0}_{B}}}^{2},
                {m_{\tilde{e}_{X}}}^{2},
                {m_{\tilde{e}_{Y}}}^{2}),
 \\
g_{1}^{c} & = & -\frac{\sqrt{2}}{64\pi^2 G_F}\sum_{A,B=1}^{2}\sum_{X,Y=1}^{3}
        C^{L}_{2AX} C^{R*}_{1AY} C^{L}_{1BY} C^{R*}_{1BX} \nonumber \\
          &   &m_{\tilde{\chi}^{-}_{A}}m_{\tilde{\chi}^{-}_{B}}
         d_{0}({m_{\tilde{\chi}^{-}_{A}}}^{2},
               {m_{\tilde{\chi}^{-}_{B}}}^{2},
               {m_{\tilde{\nu}_{X}}}^{2},
               {m_{\tilde{\nu}_{Y}}}^{2}).
\end{eqnarray}
 The coupling constant $g_{3}$ is divided into three parts. $g_{31}$ is a contribution of box diagrams and $g_{32}$ is that of Z-penguin diagrams. $g_{33}$ is a contribution of a off-shell photon-penguin diagrams. 

\begin{equation}
g_{3} = g_{31} + g_{32} + g_{33},
\end{equation}

\begin{eqnarray}
g_{31}^{n} & = & -\frac{\sqrt{2}}{64\pi^2 G_F}\sum_{A,B=1}^{4}\sum_{X,Y=1}^{6}
       \{ 
         N^{L}_{2AX} N^{L*}_{1AY} N^{L}_{1BY} N^{L*}_{1BX}
         d_{2}({m_{\tilde{\chi}^{0}_{A}}}^{2},
               {m_{\tilde{\chi}^{0}_{B}}}^{2},
               {m_{\tilde{e}_{X}}}^{2},
               {m_{\tilde{e}_{Y}}}^{2}) \nonumber \\
          &   &  +\frac{1}{2}
         N^{L}_{2AX} N^{L}_{1AY} N^{L*}_{1BY} N^{L*}_{1BX}
         m_{\tilde{\chi}^{0}_{A}}  
         m_{\tilde{\chi}^{0}_{B}}
         d_{0}({m_{\tilde{\chi}^{0}_{A}}}^{2},
               {m_{\tilde{\chi}^{0}_{B}}}^{2},
               {m_{\tilde{e}_{X}}}^{2},
               {m_{\tilde{e}_{Y}}}^{2})\},
 \\
g_{32}^{n} & = & - \frac{1}{16\pi^{2}} Z^{e}_{R}
                 [\sum_{A,B=1}^{4}\sum_{X=1}^{6}
                 N^{L}_{2AX} N^{L*}_{1BX} 
                 \{4(Y_{\tilde{\chi}^{0}_{L}})_{AB}
                 c_{2}({m_{\tilde{e}_{X}}}^{2},
                       {m_{\tilde{\chi}^{0}_{A}}}^{2},
                       {m_{\tilde{\chi}^{0}_{B}}}^{2}) \nonumber \\
           &   & -2
                            m_{\tilde{\chi}^{0}_{A}}
                            m_{\tilde{\chi}^{0}_{B}}
                   (Y_{\tilde{\chi}^{0}_{R}})_{AB}
                   c_{0}({m_{\tilde{e}_{X}}}^{2},
                         {m_{\tilde{\chi}^{0}_{A}}}^{2},
                         {m_{\tilde{\chi}^{0}_{B}}}^{2}) 
                  \} \nonumber \\
           &   & +
                 \sum_{A=1}^{4}\sum_{X,Y=1}^{6}
                 N^{L}_{2AX} N^{L*}_{1AY}
                 (X_{\tilde{e}_{L}})_{XY} 
                 c_{2}({m_{\tilde{\chi}^{0}_{A}}}^{2},
                       {m_{\tilde{e}_{X}}}^{2},
                       {m_{\tilde{e}_{Y}}}^{2}) ],
 \\
g_{33}^{n} & = & -\frac{\sqrt{2}e^{2}}{1152\pi^2 G_F}\sum_{A=1}^{4}\sum_{X=1}^{6}
                 N^{L}_{2AX} N^{L*}_{1AX}
                 \frac{1}{m_{\tilde{e}_{X}}^{2}}
                 b_{0}^{n}(\frac{m_{\tilde{\chi}^{0}_{A}}^{2}}
                                {m_{\tilde{e}_{X}}^{2}}),
 \\
g_{31}^{c} & = & -\frac{\sqrt{2}}{64\pi^2 G_F}\sum_{A,B=1}^{2}\sum_{X,Y=1}^{3}
        C^{L}_{2AX} C^{L*}_{1AY} C^{L}_{1BY} C^{L*}_{1BX} \nonumber \\
 & &    d_{2}({m_{\tilde{\chi}^{-}_{A}}}^{2},
                {m_{\tilde{\chi}^{-}_{B}}}^{2},
                {m_{\tilde{\nu}_{X}}}^{2},
                {m_{\tilde{\nu}_{Y}}}^{2}),
 \\
g_{32}^{c} & = & -\frac{1}{16\pi^{2}} Z^{e}_{R}
                  \sum_{A,B=1}^{2}\sum_{X=1}^{3}
                  C^{L}_{2AX} C^{L*}_{1BX} 
                  \{4(Y_{\tilde{\chi}^{-}_{L}})_{AB}
                 c_{2}({m_{\tilde{\nu}_{X}}}^{2},
                       {m_{\tilde{\chi}^{-}_{A}}}^{2},
                       {m_{\tilde{\chi}^{-}_{B}}}^{2}) \nonumber \\
           &   & -2
                        m_{\tilde{\chi}^{-}_{A}}
                        m_{\tilde{\chi}^{-}_{B}}
                    (Y_{\tilde{\chi}^{-}_{R}})_{AB}
                   c_{0}({m_{\tilde{\nu}_{X}}}^{2},
                         {m_{\tilde{\chi}^{-}_{A}}}^{2},
                         {m_{\tilde{\chi}^{-}_{B}}}^{2})\},
 \\
g_{33}^{c} & = &  -\frac{\sqrt{2} e^2}{1152\pi^2 G_F}\sum_{A=1}^{2}\sum_{X=1}^{3}
                  C^{L}_{2AX} C^{L*}_{1AX}
                  \frac{1}{m_{\tilde{\nu}_{X}}^{2}}
                  b_{0}^{c}(\frac{m_{\tilde{\chi}^{-}_{A}}^{2}}
                                 {m_{\tilde{\nu}_{X}}^{2}}).
\end{eqnarray}
 The coupling constant $g_{5}$ is divided into three parts. $g_{51}$ is a contribution of box diagrams and $g_{52}$ is that of Z-penguin diagrams. $g_{53}$ is a contribution of a off-shell photon-penguin diagrams. 

\begin{equation}
g_{5} = g_{51} + g_{52} + g_{53},
\end{equation}

\begin{eqnarray}
g_{51}^{n} & = & -\frac{\sqrt{2}}{64\pi^{2}G_F}\sum_{A,B=1}^{4}\sum_{X,Y=1}^{6}
                 \{
                 (N^{L}_{2AX} N^{L*}_{1AY} N^{R}_{1BY}  N^{R*}_{1BX}
                 -N^{L}_{2AX} N^{R}_{1AY}  N^{R*}_{1BY} N^{L*}_{1BX} \nonumber \\
 & &             +N^{L}_{2AX} N^{R}_{1AY}  N^{L*}_{1BY} N^{R*}_{1BX})
                 d_{2}({m_{\tilde{\chi}^{0}_{A}}}^{2},
                       {m_{\tilde{\chi}^{0}_{B}}}^{2},
                       {m_{\tilde{e}_{X}}}^{2},
                       {m_{\tilde{e}_{Y}}}^{2}) \nonumber \\
 & &              -\frac{1}{2}m_{\tilde{\chi}^{0}_{A}}
                             m_{\tilde{\chi}^{0}_{B}}
                  N^{L}_{2AX}N^{R*}_{1AY}N^{R}_{1BY}N^{L*}_{1BX}
                  d_{0}({m_{\tilde{\chi}^{0}_{A}}}^{2},
                        {m_{\tilde{\chi}^{0}_{B}}}^{2},
                        {m_{\tilde{e}_{X}}}^{2},
                        {m_{\tilde{e}_{Y}}}^{2})
                 \},
 \\
g_{52}^{n} & = & -\frac{1}{16\pi^{2}} Z^{e}_{L}
                  [\sum_{A,B=1}^{4}\sum_{X=1}^{6}
                  N^{L}_{2AX} N^{L*}_{1BX}
                  \{4(Y_{\tilde{\chi}^{0}_{L}})_{AB}
                   c_{2}({m_{\tilde{e}_{X}}}^{2},
                         {m_{\tilde{\chi}^{0}_{A}}}^{2},
                         {m_{\tilde{\chi}^{0}_{B}}}^{2}) \nonumber \\
           &   &  -2
                             m_{\tilde{\chi}^{0}_{A}}
                             m_{\tilde{\chi}^{0}_{B}}
                  (Y_{\tilde{\chi}^{0}_{R}})_{AB}
                   c_{0}({m_{\tilde{e}_{X}}}^{2},
                         {m_{\tilde{\chi}^{0}_{A}}}^{2},
                         {m_{\tilde{\chi}^{0}_{B}}}^{2}) 
                  \} \nonumber \\
           &   & +
                 \sum_{A=1}^{4}\sum_{X,Y=1}^{6}
                 N^{L}_{2AX} N^{L*}_{1AY}
                 (X_{\tilde{e}_{L}})_{XY} 
                 c_{2}({m_{\tilde{\chi}^{0}_{A}}}^{2},
                       {m_{\tilde{e}_{X}}}^{2},
                       {m_{\tilde{e}_{Y}}}^{2}) ],
 \\
g_{53}^{n} & = & g_{33}^{n},
 \\
g_{51}^{c} & = & -\frac{\sqrt{2}}{64\pi^2 G_F}\sum_{A,B=1}^{2}\sum_{X,Y=1}^{3}
                 \{C^{L}_{2AX} C^{L*}_{1AY} C^{R}_{1BY} C^{R*}_{1BX}
                 d_{2}({m_{\tilde{\chi}^{-}_{A}}}^{2},
                       {m_{\tilde{\chi}^{-}_{B}}}^{2},
                       {m_{\tilde{\nu}_{X}}}^{2},
                       {m_{\tilde{\nu}_{Y}}}^{2}) \nonumber \\
           &   &-\frac{1}{2}
                   C^{L}_{2AX} C^{R*}_{1AY} C^{R}_{1BY} C^{L*}_{1BX}
                   m_{\tilde{\chi}^{-}_{A}}
                   m_{\tilde{\chi}^{-}_{B}}
                 d_{0}({m_{\tilde{\chi}^{-}_{A}}}^{2},
                       {m_{\tilde{\chi}^{-}_{B}}}^{2},
                       {m_{\tilde{\nu}_{X}}}^{2},
                       {m_{\tilde{\nu}_{Y}}}^{2})
         \},
 \\ 
g_{52}^{c} & = & -\frac{1}{16\pi^{2}} Z^{e}_{L}
                 \sum_{A,B=1}^{2}\sum_{X=1}^{3}
                 C^{L}_{2AX} C^{L*}_{1BX}
                  \{4(Y_{\tilde{\chi}^{-}_{L}})_{AB}
                 c_{2}({m_{\tilde{\nu}_{X}}}^{2},
                       {m_{\tilde{\chi}^{-}_{A}}}^{2},
                       {m_{\tilde{\chi}^{-}_{B}}}^{2}) \nonumber \\
           &   & -2
                       m_{\tilde{\chi}^{-}_{A}}
                       m_{\tilde{\chi}^{-}_{B}}
                   (Y_{\tilde{\chi}^{-}_{R}})_{AB}
                 c_{0}({m_{\tilde{\nu}_{X}}}^{2},
                       {m_{\tilde{\chi}^{-}_{A}}}^{2},
                       {m_{\tilde{\chi}^{-}_{B}}}^{2})\},
 \\
g_{53}^{c} & = & g_{33}^{c}.
\end{eqnarray}
 Various mixing matrices and Z coupling constants which appear in the above formulas are given as follows:

\begin{eqnarray}
(Y_{\tilde{\chi}^{0}_{L}})_{AB} & = & -\frac{1}{2}
                 \{ (O_{NL})_{A3}(O_{NL})_{B3}^{*}-
                    (O_{NL})_{A4}(O_{NL})_{B4}^{*} 
                 \},
 \\
(Y_{\tilde{\chi}^{0}_{R}})_{AB} & = & \frac{1}{2}
                 \{ (O_{NR})_{A3}(O_{NR})_{B3}^{*}-
                    (O_{NR})_{A4}(O_{NR})_{B4}^{*} 
                 \},
\end{eqnarray}

\begin{eqnarray}
(Y_{\tilde{\chi}^{-}_{L}})_{AB} & = & -\frac{1}{2}
                                      (O_{CL})_{A2} (O_{CL})_{B2}^{*},
 \\
(Y_{\tilde{\chi}^{-}_{R}})_{AB} & = & -\frac{1}{2}
                                      (O_{CR})_{A2} (O_{CR})_{B2}^{*},
\end{eqnarray}

\begin{eqnarray}
(X_{\tilde{e}_{L}})_{XY} & = & -\sum_{k=1}^{3}(U_{e})_{Xk}(U_{e})^{*}_{Yk},
 \\
(X_{\tilde{e}_{R}})_{XY} & = & \sum_{k=1}^{3}(U_{e})_{Xk+3}(U_{e})^{*}_{Yk+3}.
\end{eqnarray}

\begin{eqnarray}
Z^{e}_{L} & = & (-\frac{1}{2}+\sin^{2}\theta_{W}),
 \\
Z^{e}_{R} & = & \sin^{2}\theta_{W}.
\end{eqnarray}
The photon-penguin coupling constant is written as follows:

\begin{equation}
A_{R} = A_{R}^{n}+A_{R}^{c},
\end{equation}

\begin{eqnarray}
A_{R}^{n} & = & \frac{\sqrt{2}e}{256\pi^2 G_F}
          \sum_{A=1}^{4}\sum_{X=1}^{6}\frac{1}{m_{\tilde{e}_{X}}^{2}}
          \{ \frac{1}{6}N^{R}_{2AX} N^{R*}_{1AX} 
            b_{1}^{n}(\frac{m_{\tilde{\chi}^{0}_{A}}^{2}}
                           {m_{\tilde{e}_{X}}^{2}}) \nonumber \\
     & &   + N^{L}_{2AX} N^{R*}_{1AX} 
            \frac{m_{\tilde{\chi}^{0}_{A}}}{m_{\mu}}
            b_{2}^{n}(\frac{m_{\tilde{\chi}^{0}_{A}}^{2}}
                           {m_{\tilde{e}_{X}}^{2}}) \},
 \\
A_{R}^{c} & = & -\frac{\sqrt{2}e}{128\pi^2 G_F}
          \sum_{A=1}^{2}\sum_{X=1}^{3}\frac{1}{m_{\tilde{\nu}_{X}}^{2}}
          \{ \frac{1}{6}C^{R}_{2AX} C^{R*}_{1AX} 
          b_{1}^{c}(\frac{m_{\tilde{\chi}^{-}_{A}}^{2}}
                         {m_{\tilde{\nu}_{X}}^{2}}) \nonumber \\
          &   &+ C^{L}_{2AX} C^{R*}_{1AX}
            \frac{m_{\tilde{\chi}^{-}_{A}}}{m_{\mu}}
            b_{2}^{c}(\frac{m_{\tilde{\chi}^{-}_{A}}^{2}}
                           {m_{\tilde{\nu}_{X}}^{2}}) \}.
\end{eqnarray}
 The other coupling constants can be obtained by simply exchanging the suffix of above formulas:

\begin{eqnarray}
g_{2} &=& g_{1}(L \leftrightarrow R),
 \\
g_{4} &=& g_{3}(L \leftrightarrow R),
 \\
g_{6} &=& g_{5}(L \leftrightarrow R),
 \\
A_{L} &=& A_{R}(L \leftrightarrow R).
\end{eqnarray}

\subsection{Mass Functions}

 The mass functions used in the effective coupling constants of the $\mu^{+}\rightarrow e^{+}\gamma$ and $\mu^{+} \rightarrow e^{+}e^{+}e^{-}$ processes are defined as follows:

\begin{eqnarray}
b_{0}^{n}(x) & = &\frac{1}{2(1-x)^{4}}(2-9x+18x^{2}-11x^{3}+6x^{3}\ln(x)), 
 \\
b_{1}^{n}(x) & = &\frac{1}{(1-x)^{4}}(1-6x+3x^{2}+2x^{3}-6x^{2}\ln(x)), 
 \\
b_{2}^{n}(x) & = &\frac{1}{(1-x)^{3}}(1-x^{2}+2x\ln(x)),
 \\
b_{0}^{c}(x) & = & \frac{1}{2(1-x)^{4}}(-16+45x-36x^{2}+7x^{3}+6(3x-2)\ln(x)),
 \\
b_{1}^{c}(x) & = & \frac{1}{2(1-x)^{4}}(2+3x-6x^{2}+x^{3}+6x\ln(x)),
 \\
b_{2}^{c}(x) & = & \frac{1}{2(1-x)^{3}}(-3+4x-x^{2}-2\ln(x)),
\end{eqnarray}

\begin{eqnarray}
c_{0}(x,y,z) & = & -\frac{x\ln(x)}{(y-x)(z-x)}-\frac{y\ln(y)}{(x-y)(z-y)}
                -\frac{z\ln(z)}{(x-z)(y-z)},
 \\
c_{2}(x,y,z) & = & \frac{1}{4}[\frac{3}{2}-\frac{x^{2}\ln(x)}{(y-x)(z-x)}
                -\frac{y^{2}\ln(y)}{(x-y)(z-y)}
                -\frac{z^{2}\ln(z)}{(x-z)(y-z)}
                 ],\nonumber \\
\end{eqnarray}

\begin{eqnarray}
d_{0}(x,y,z,w) & = &    \frac{x\ln(x)}{(y-x)(z-x)(w-x)} 
                       +\frac{y\ln(y)}{(x-y)(z-y)(w-y)} \nonumber \\
               &   &   +\frac{z\ln(z)}{(x-z)(y-z)(w-z)}
                       +\frac{w\ln(w)}{(x-w)(y-w)(z-w)},
 \\
d_{2}(x,y,z,w) & = &   \frac{1}{4} \{ \frac{x^{2}\ln(x)}{(y-x)(z-x)(w-x)} 
                       +\frac{y^{2}\ln(y)}{(x-y)(z-y)(w-y)} \nonumber \\
               &   &   +\frac{z^{2}\ln(z)}{(x-z)(y-z)(w-z)}
                       +\frac{w^{2}\ln(w)}{(x-w)(y-w)(z-w)} \}.
\end{eqnarray}

\section{Neutron EDM}
\label{sec:nedm}
We discuss QCD correction in the calculation of the neutron EDM \cite{QCD Correction},\cite{98Nath}.
The neutron EDM are calculated by the following effective Lagrangian.
\begin{eqnarray}
  \label{eq:Heff}
  {\cal L}_{\rm eff} = \sum_{q}~C^{E}_q (\mu) {\cal O}^E_q(\mu) 
  + \sum_{q}~C_q^C(\mu) {\cal O}^C_q(\mu)
  + C^G(\mu) {\cal O}^G(\mu),
\end{eqnarray}
where ${\cal O}^E_q$, ${\cal O}^C_q$, ${\cal O}^G$ correspond to the quark
electric dipole, chromomagnetic dipole, and gluonic Weinberg's operators,
respectively, which are given by
\begin{eqnarray}
  {\cal O}^{E}_q &=& -\frac{i}{2} \overline{q} \sigma_{\mu\nu} \gamma_5
  q F^{\mu\nu},
\\
  {\cal O}^{C}_q &=& -\frac{i}{2} \overline{q} \sigma_{\mu\nu} \gamma_5 T^a
  q G^{a\mu\nu},
\\
  {\cal O}^G  &=&  -\frac{1}{6} f^{abc} \epsilon^{\mu\nu\lambda\rho}
  G^a_{\lambda\rho} G^{b}_{\mu\alpha} G^{c\alpha}_{\nu}.
\end{eqnarray}
Here,
$\epsilon^{0123}=1$, and $f^{abc}$ is the structure constant of the
SU(3) group. 

In SUSY models, we can obtain the Wilson coefficients at the electroweak scale by
evaluating 1-loop diagrams.  $C^E_q$ is induced by the photon-penguin
diagram.
$C^{E(\tilde{\chi}^-)}_q $ and $C^{E(\tilde{\chi}^0)}_q$ are induced by the photon- and gluino-penguin diagrams.
There are three types of SUSY contribution, chargino-squark loop, neutralino-squark loop and gluino-squark loop diagrams.
The gluonic Weinberg's operator is induced at a 2-loop
level and the diagram involving the stop and the gluino gives dominant
contribution. 
These contributions are listed in Ref.\cite{98Nath}.

We can take into account a QCD correction from the electroweak scale
to  a hadronic scale (1 GeV), by using 
the following renormalization group equations for the Wilson coefficients.
\begin{eqnarray}
  \label{eq:rge}
  \mu \frac{d \vec{C}(\mu)}{d \mu} = \frac{\alpha_s(\mu)}{4\pi} \gamma^T
  \vec{C}(\mu),  
\end{eqnarray}
where $\vec{C} = (C^E_q, C^C_q, C^G)^T$ and the anomalous dimension
matrix $\gamma_{ij}$ is written by 
\begin{eqnarray}
  \label{eq:gamma}
  \gamma =
  \left(
    \begin{array}{ccc}
       8/3 & 0 &0\\
       32eQ/(3g_s)  &   (-29 + 2 N_f)/3  & 0\\
       0           &  6 m_q &  2 N_f + 3\\
    \end{array}
    \right).
\end{eqnarray}
Here, $N_f$ is a number of the quark flavor and $Q$ denotes the
electro-magnetic charge of the quark in unit $e$ $(e >0)$.
The RGEs can be solved analytically as follows:
\begin{eqnarray}
  \label{eq:solution}
C_q^E(\mu) &=& \eta^{\frac{8}{33-2N_f}} \left[C^E_q(\mu_{0})
+8eQ \left(1-\eta^{\frac{-4}{33-2N_f}}\right) \frac{C^C_q(\mu_{0})}{g_s(\mu_{0})}
\right.
\nonumber \\
&-&
\left.
\frac{72eQm_q(\mu_{0})}{7+2N_f}
\left(1- \eta^{\frac{-4}{33-2N_f}} + \frac{2}{2N_f+5}
 (1- \eta^{\frac{10+4N_f}{33-2N_f}}) \right)\frac{C_G(\mu_{0})}{g_s(\mu_{0})}
\right],~~~~~
\\
C_q^C(\mu) &=& \eta^{\frac{-29+2 N_f}{33-2N_f}} \left[C^C_q (\mu_{0})
\right.
\nonumber \\
&&
\left.- \frac{9}{7+2N_f} \left(1-\eta^{\frac{14+4N_f}{33-2N_f}} \right)
m_q(\mu_{0}) C^G(\mu_{0})\right],
\\
C^G(\mu) &=& \eta^{\frac{9 +6 N_f }{33 - 2 N_f}} C^G(\mu_{0}),
\end{eqnarray}
where $\eta=g_s (\mu_{0})/ g_s (\mu)$.

We solve RGE from $m_{W}$ to $m_{b}$, $m_{b}$ to $m_{c}$ and $m_{c}$ to the 1 GeV scale.
When the heavy quarks $(c, b)$ decouple at their
mass threshold, $C^G$ is induced through the chromo-electric dipole
moment of the heavy quarks.
Difference $C^{G}$ below and above the threshold is given by \cite{threshold correction}  
\begin{eqnarray}
  \label{eq:threshold}
  C^G( m_q )_{below} - C^G( m_q )_{above} = 
  + \frac{\alpha_s(m_q)}{8\pi m_q(m_q)} C^C_q(m_q).
\end{eqnarray}
Taking into account the QCD and threshold corrections, we obtain the
effective Lagrangian at the hadronic scale.
It is then straightforward to evaluate the effective $\cal L \mit$ at 1 GeV scale from $m_{W}$ scale.

The neutron EDM $(d_n)$ is given by the Wilson coefficients at a
hadronic scale as follows:
\begin{eqnarray}
  \label{eq:nEDM}
  d_n &=& d_n^E + d_n^C + d_n^G,
  \\
  d_n^E &=& \frac{1}{3}\left(4 C_d^E - C_u^E\right),
    \\
    d_n^C &=& \frac{1}{3}\frac{e}{4\pi}\left(4 C_d^C - C_u^C\right),
    \\
    d_n^G &=& \frac{eM}{4\pi} C^G,
\end{eqnarray}
where $M$ is a chiral symmetry breaking parameter, which is estimated
as $1.19$ GeV. 
In the above we use non-relativistic quark model for $d_{n}^{E}$ and naive dimensional analysis for $d_{n}^{C}$ and $d_{n}^{G}$.

\newpage

\hspace{-1.2cm}{\Large {\bf Figure Captions:}}

\begin{figure}[htbp]
  \begin{center}
    \leavevmode
    \scalebox{.8}{\includegraphics{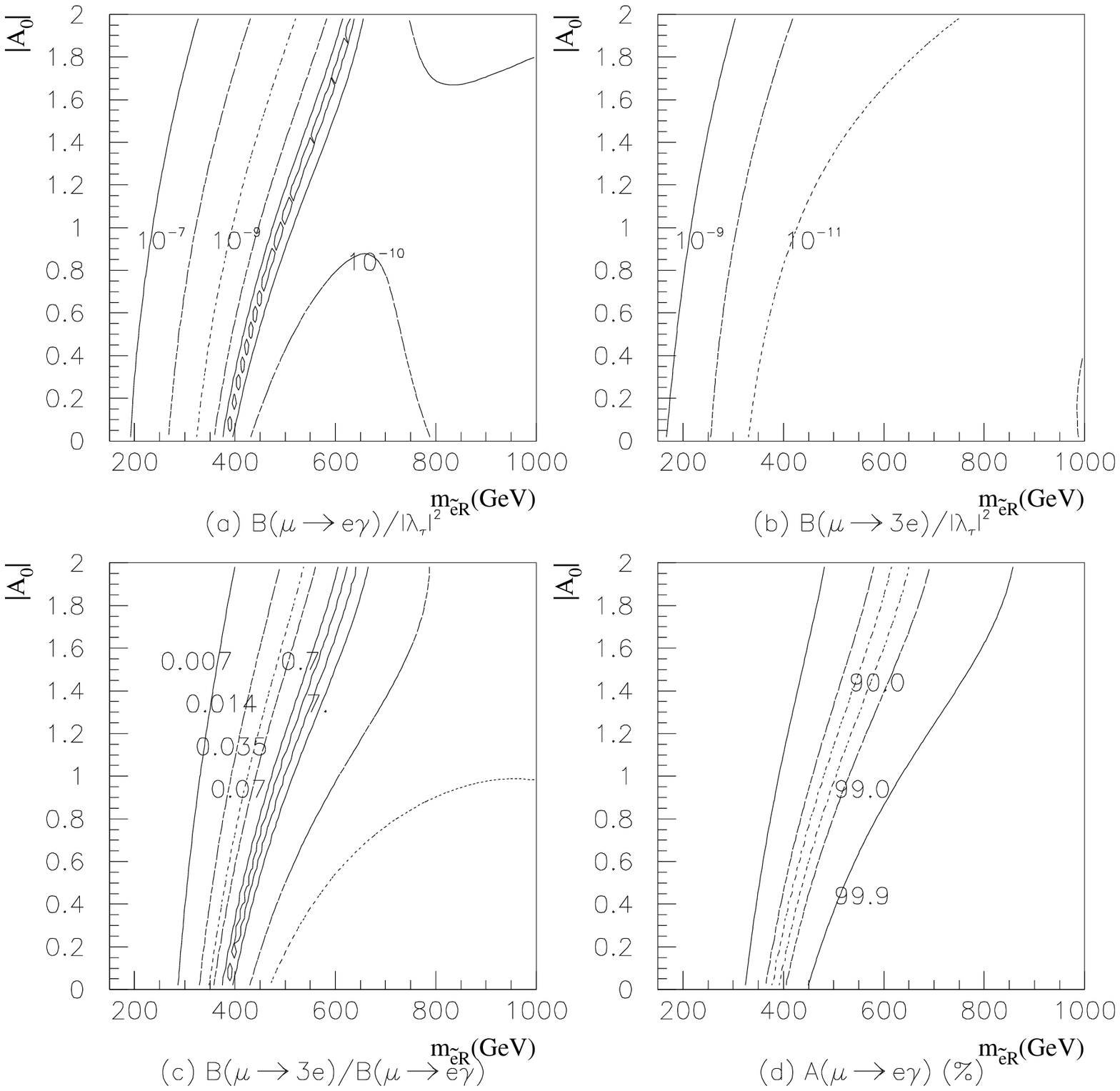}}
  \end{center}
\end{figure}

\begin{figure}[htbp]
  \begin{center}
    \leavevmode
    \scalebox{.8}{\includegraphics{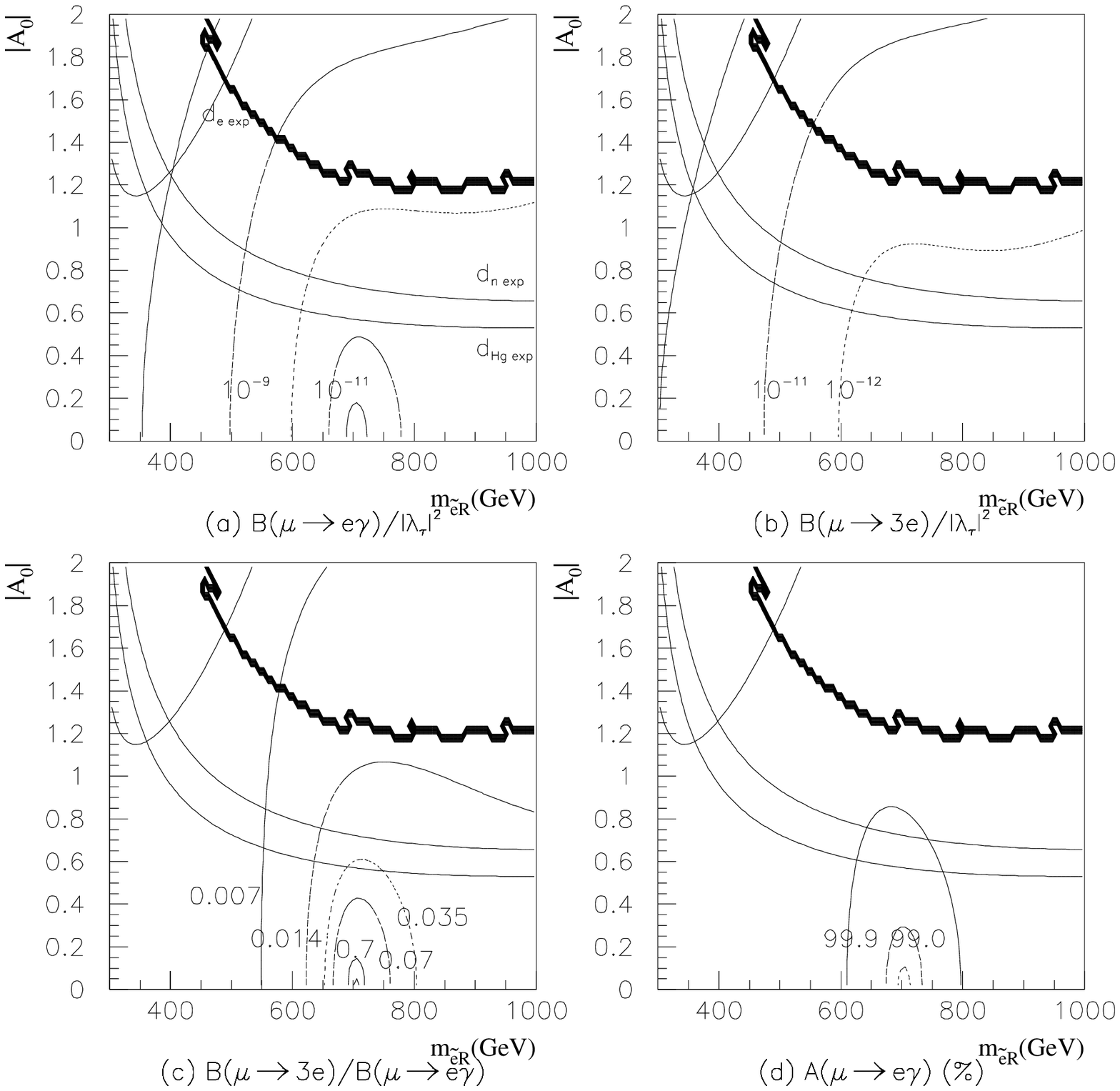}}
  \end{center}
\end{figure}

\begin{figure}[htbp]
  \begin{center}
    \leavevmode
    \scalebox{.8}{\includegraphics{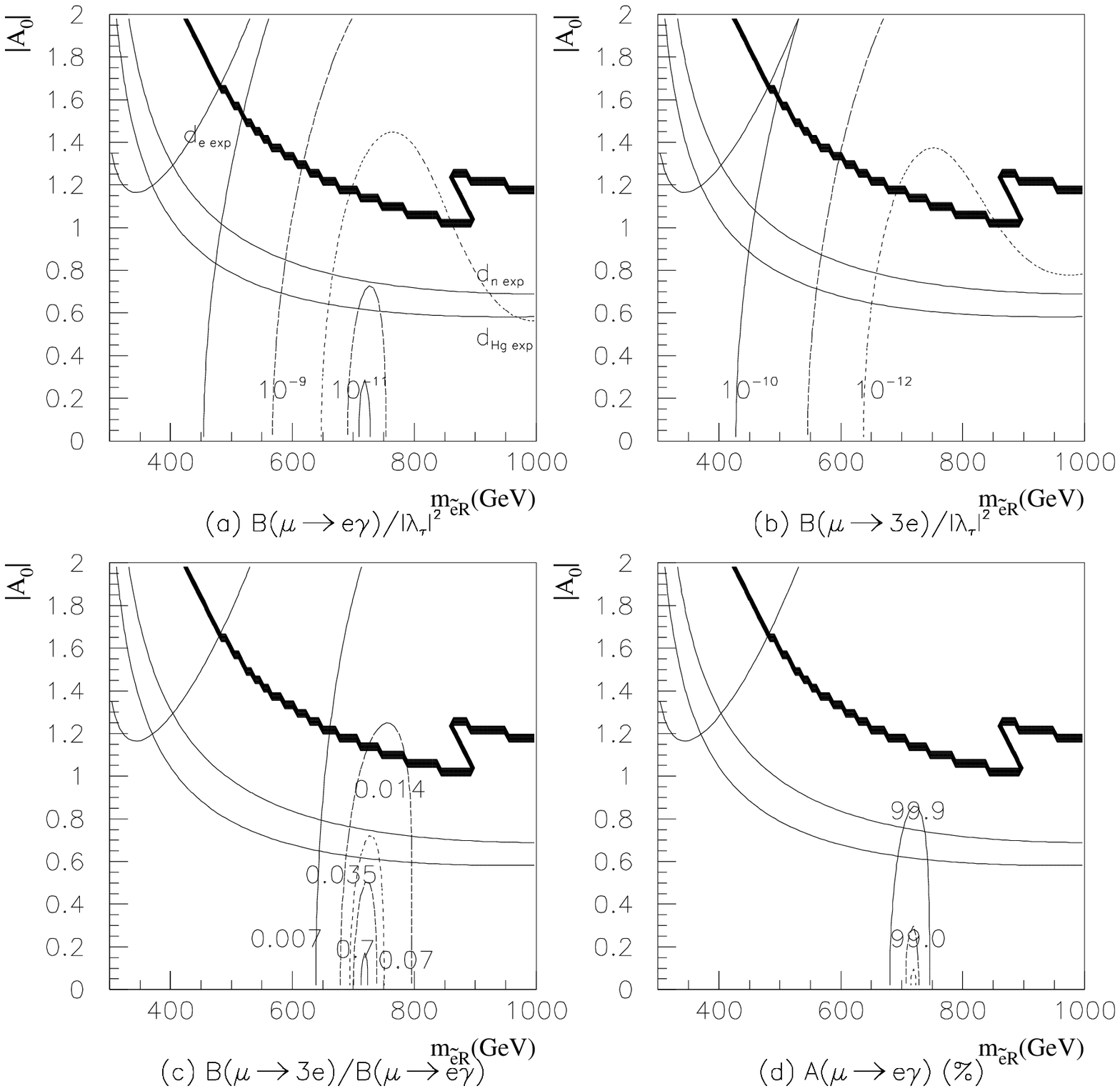}}
  \end{center}
\end{figure}

\begin{figure}[htbp]
  \begin{center}
    \leavevmode
    \scalebox{.8}{\includegraphics{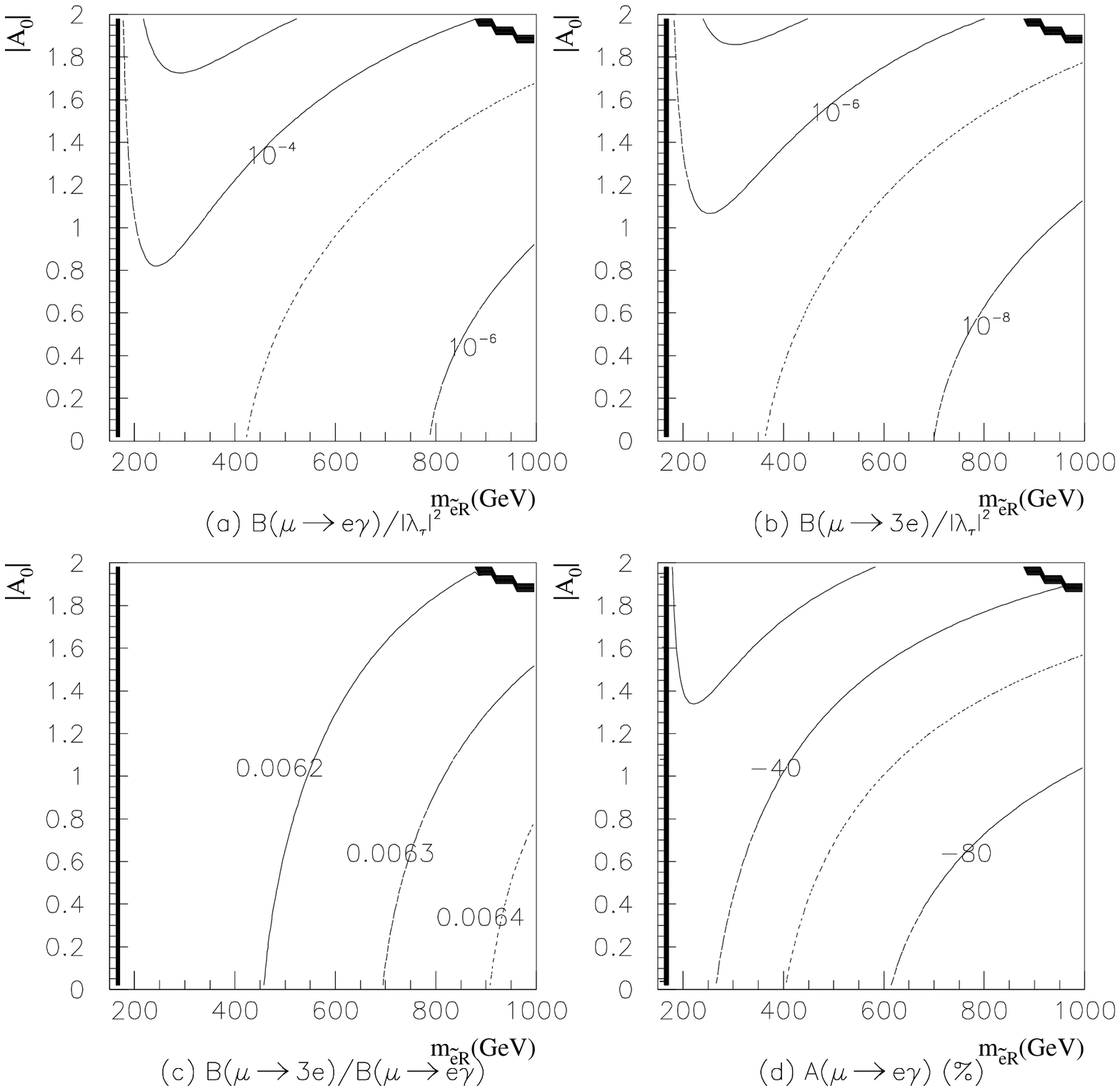}}
  \end{center}
\end{figure}

\begin{figure}[htbp]
  \begin{center}
    \leavevmode
    \scalebox{.8}{\includegraphics{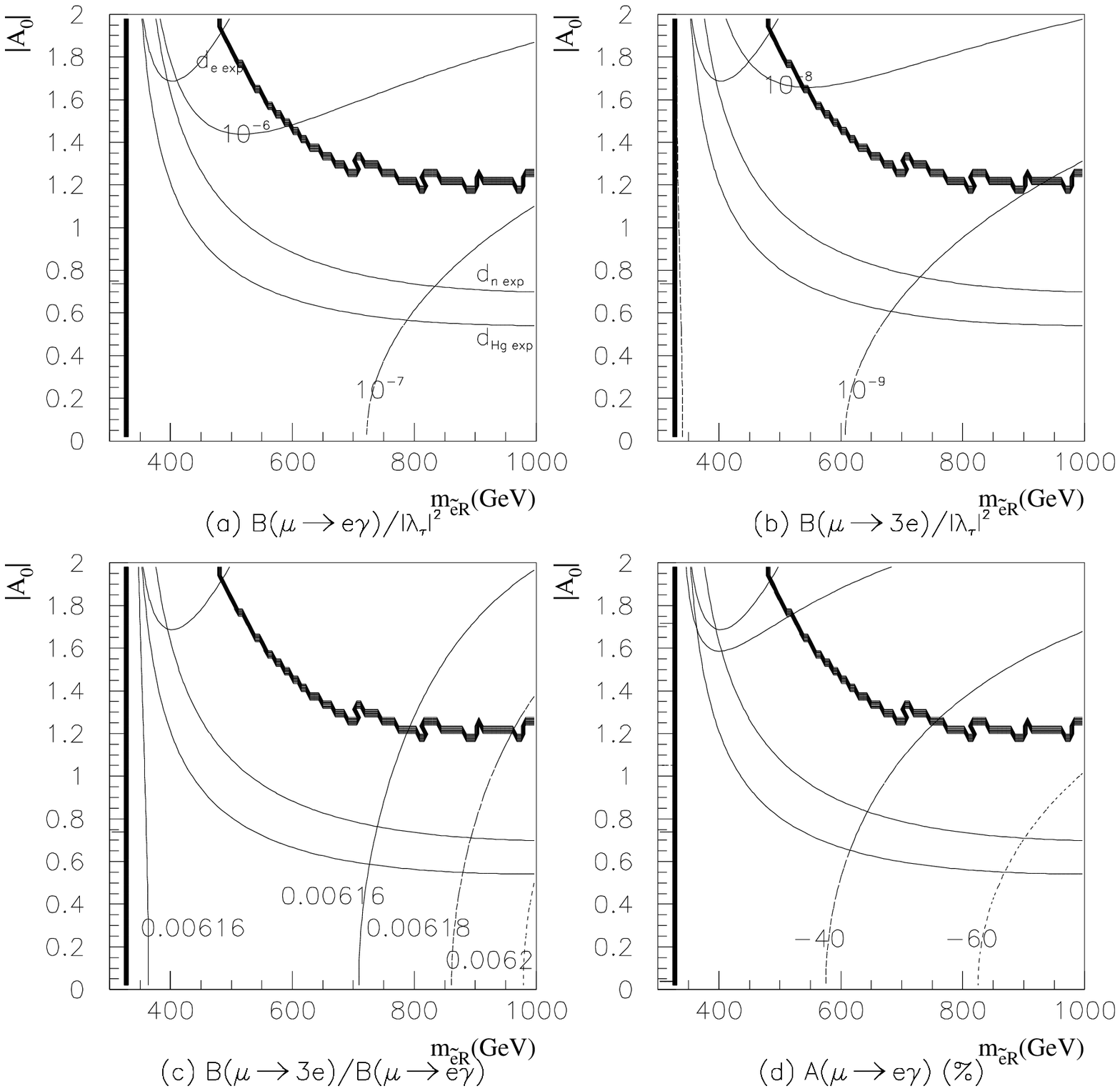}}
  \end{center}
\end{figure}

\end{document}